\documentclass{jpp}
\usepackage{graphicx}
\usepackage{epsfig}
\usepackage{hyperref}
\usepackage{amsmath}
\usepackage{xcolor}

\def\BEQ {\begin{equation}} 
\def\EEQ {\end{equation}}
\def\BEQN {\begin{displaymath}} 
\def\EEQN {\end{displaymath}}
\def\BEQAR {\begin{eqnarray}}
\def\EEQAR {\end{eqnarray}}
\def\non {\nonumber}

\def\En#1{\label{Eq.#1}}
\def\E#1{(\ref{Eq.#1})}

\def\R#1{\citep{#1}}
\def\Ron#1{\citet{#1}}
\def\Fn#1{\label{Fig.#1}}
\def\F#1{\ref{Fig.#1}}
\def\T#1{\ref{Table #1}}
\def\Tn#1{\label{Table #1}}
\def\Sn#1{\label{Sec.#1}}
\def\S#1{\ref{Sec.#1}}

\def\hor #1{{\hskip #1 mm}}
\def\ver #1{{\vskip #1 mm}}
\def\hs {{\hskip 0.2mm}}

\def\bfe {{\bf e}}

\def\bfj {{\bf j}}
\def\bfk {{\bf k}}
\def\bfn {{\bf n}}

\def\bfv {{\bf v}}
\def\bfB {{\bf B}}
\def\bfD {{\bf D}}
\def\bfE {{\bf E}}
\def\bfF {{\bf F}}

\def\bfH {{\bf H}}
\def\bfQ {{\bf Q}}
\def\bfR {{\bf R}}
\def\bfS {{\bf S}}

\def\bfxi{\mbox{\boldmath $\xi$}}  

\def\ddx #1{\frac{d #1}{dx}}
\def\ddxx #1{\frac{d^2 #1}{dx^2}}

\def\deldelt #1{\frac{\partial #1}{\partial t}}
\def\deldeltt #1{\frac{\partial^2 #1}{\partial t^2}}
\def\deldelxx #1{\frac{\partial^2 #1}{\partial x^2}}
\def\ddxx #1{\frac{d^2 #1}{d x^2}}

\def\div #1{\nabla\cdot #1}
\def\curl #1{\nabla\times #1}

\def\oms {{\omega^2}}

\def\IT #1{{\it #1\/}}

\def\ds{\displaystyle}

\def\half {{\textstyle\frac{1}{2}}}
\def\onefourth {{\textstyle\frac{1}{4}}}

\def\doublel {\lbrack\!\lbrack}
\def\doubler {\rbrack\!\rbrack}

\shorttitle{Once more: Leaky MHD modes}
\shortauthor{Hans Goedbloed \& Rony Keppens}

\title{Once more: Leaky MHD waves in coronal magnetic flux tubes.\\[2mm]{\em\Large Analogous to leaky EM waves in dielectric media?}}

\author{Hans Goedbloed\aff{1}
  \corresp{\email{goedbloed@differ.nl}}
 \and Rony Keppens\aff{2}}

\affiliation{\aff{1}DIFFER -- Dutch Institute for Fundamental Enery 
Research, de Zaale 20, 5612 AJ Eindhoven, the Netherlands
\aff{2}Centre for mathematical Plasma Astrophysics, KU Leuven, Belgium}

\begin{document}

\maketitle

\begin{abstract}
{\bf Abstract} 
By a detailed comparison of  leaky magnetohydrodynamic waves in coronal magnetic flux tubes with leaky electromagnetic waves in dielectric media it is shown that 
the latter kind may be called quasi-normal modes, since they can be regularised by a normalisation which systematically cuts off the contribution of the external homogeneous region, whereas such a possibility is forbidden for the former kind by the conservation of magnetic flux. Consequently, leaky magnetohydrodynamic waves cannot be systematically applied to coronal seismology, i.e.\ to the inverse spectral problem of determining the different equilibrium distributions of the fields by  comparing the spectra they produce with  the observed ones.
\end{abstract}

\section{Introduction}{\Sn{1}}

In our recent paper, \Ron{GKP2023}, we discarded `leaky modes in coronal magnetic flux 
tubes' as a possible physical model to describe observable effects of magnetohydrodynamic (MHD) 
waves in the solar corona. Consequently, we also had to conclude that the corresponding text 
in Section~10.5 of our textbook~\R{GKP2019}  should be deleted. At about the same time, 
an extensive review paper appeared by~\Ron{Huang2023} on leaky electromagnetic (EM) waves in dielectric media. 
In Appendix~A of that paper the explicit  dispersion equations are given for leaky modes in 
dielectric structures due to a discontinuity of the phase velocity. Those are identical to the 
corresponding expressions for leaky modes in coronal flux tubes. Since the mathematical 
expressions are identical for the two cases of EM waves propagating in a 
dielectric slab and MHD waves in a corresponding magnetised plasma slab, and the evidence of 
the reality of leaky modes in dielectrics is overwhelming, the obvious question to be answered 
is whether the physical consequences are then also the same. In particular, do we have to 
retract our previous conclusion on the absence of a physical meaning of leaky modes in 
coronal flux tubes? In this paper, we will address this issue. 

We will analyse the two cases of EM waves in dielectric media (Secs.~\S{2}--\S{4}) and 
MHD waves in coronal magnetic flux tubes (Secs.~\S{5}--\S{7}), side-by-side, for the simple configuration 
depicted in Fig.\F{1}. We will draw conclusions on the possible existence of leaky modes in coronal flux tubes in Sec.~\S{8}.


 \ver{10}

\section{Electrodynamic waves in a dielectric slab}{\Sn{2}} 

\subsection{Basic equations}{\Sn{2.1}}

We exploit Maxwell's equations in the absence of currents and space charges, where 
one-dimensional inhomogeneity occurs through the dielectric constant, which is not constant 
though in the dielectric slab, $\epsilon = \epsilon(x)$ for $|x| \le 1$ (see Fig.~\F{1}). In this figure, the 
coordinate~$x$ has already been scaled as indicated in Eq.~\E{2.1} below. Outside the slab, 
we make the rather accurate approximation (even for air) that the magnetic permeability 
$\mu_0$ and the dielectric constant~$\epsilon_0$ have their vacuum values, 
$(\epsilon_0\mu_0)^{-1/2} = c$. This permits us to eliminate these constants of nature by 
transforming to natural units, where all space and time variables and, hence, also the phase 
velocity of the electromagnetic waves, become dimensionless and of the order unity: 
\BEQ x/a \rightarrow x \,,\qquad t\hs(c/a) \rightarrow t \,, \qquad v/c \rightarrow v 
\,,\En{2.1}\EEQ
with similar expressions for the other spatial directions. Hence, for our analysis, the only 
function of interest is the phase velocity, related to the index of refraction, $v(x) = 1/n(x)$, 
both of which only deviate from unity in the dielectric slab:
\BEQ v(x) \, \left\{ \begin{array}{l}
\le 1 \\[2mm] 
= 1 
\end{array} \right.\,,\quad n(x)
\, \left\{ \begin{array}{l}
\ge 1 \quad\quad\hbox\IT{(inside the slab, $|x| \le 1$)} \\[2mm] 
= 1 \quad\quad\hbox\IT{(outside the slab, $|x| > 1$)} 
\end{array} \right.. \En{2.2}\EEQ
The explicit functional dependence for  $v(x)$ will be chosen in Sec.~\S{3.1} when the eigenvalue spectra of 
the waves are calculated. Notice that the dielectric is assumed to be non-magnetic and also non-dispersive, i.e.\ the phase velocity does not depend on the frequency of the waves. 

In natural units, $\bfD \equiv \bfE/v^2$ and $\bfB \equiv \bfH$, so that $\bfE$ and $\bfH$ become the basic electrodynamic variables satisfying Maxwell's equations in the absence of currents and space charges:
\BEQ \frac{1}{v^2}\deldelt{\bfE}  = \curl{\bfH} \,,\quad \div{(\bfE/v^2)} = 0 \,,\qquad \deldelt{\bfH}  = - \curl{\bfE} \,,\quad \div{\bfH} = 0 \,.\En{2.3}\EEQ
For harmonic time dependence $\exp{(-{\rm{i}}\omega t})$, they  may be written in terms of the electric field alone,
\BEQ \frac{1}{v^2}\deldeltt{\bfE} + \curl{\curl{\bfE}} = 0 \quad\Rightarrow\quad 
(\oms/v^2){\bfE} - \curl{\curl{\bfE}} = 0\,.\En{2.4}\EEQ
Since $\div{(\bfE/v^2)}$ vanishes everywhere, one 
component of the electric field may be eliminated.  In this representation, the evolution of the magnetic field is just a consequence of the dynamics of the electric field:
\BEQ \deldelt{\bfH}  = - \curl{\bfE} \quad\Rightarrow\quad {\bfH}  = - 
(\rm{i}/\omega)\curl{\bfE}\,.\En{2.5}\EEQ
Hence, the problem has been reduced to solving the basic differential equation~\E{2.4} for 
two components of $\bfE$, whereas the remaining component and the three components 
of~$\bfH$ follow by algebraic expressions. Of course, one 
could also depart from the analogous equations for the magnetic field variable $\bfH$, so that the electric field variable would become secondary. 
Notice though that the differential equations for $\bfH$ differ in their dependence on $v(x)$ 
from the differential equations for $\bfE$ so that there is no symmetry between these 
variables in the dielectric.

\subsection{Ordinary differential equations for electrodynamic waves}{\Sn{2.2}}

The one-dimensional configuration depicted in Fig.~\F{1} admits a representation of the components of $\bfE$ and $\bfH$ in terms of normal modes of the form
\BEQ f(x)\hs{\rm e}^{\hs\rm{i}(\ell y + k z - \omega t)} ,\En{2.6}\EEQ
where the functional dependence on the coordinate $x$ is to be determined by the solution 
of the ordinary differential equations (ODEs) corresponding to Eqs.~\E{2.4}. The 
mode number $\ell$ in the vertical direction will be dropped right away since a rotation about the $x$-axis produces a coordinate frame with $\ell = 0\hs$. It is here introduced for 
the sake of comparison with the MHD problem, where this symmetry is not present. 
Also, in the dielectric problem for the more relevant cylindrical configuration, the translational 
symmetry in $y$ is to be replaced by the rotational symmetry in the angle~$\theta$ with the 
corresponding mode number $m$, which also cannot be transformed away. Without the mode 
number $\ell$, the ODEs~\E{2.4} become
\BEQ
\left(\begin{array}{@{}c@{\quad}c@{\quad}c@{}}
\omega^2/v^2 - k^2 & 0 & - {\rm i} \hs k \partial_x \\[2mm]
0 & \partial_x^2 + \omega^2/v^2 - k^2 & 0 \\[2mm]
- {\rm i} \hs k \partial_x & 0 &\partial_x^2 + \omega^2/v^2
\end{array} \right) 
\left(\begin{array}{@{}c}
E_x \\[2mm]
E_y \\[2mm]
E_z 
\end{array} \!\!\!\right) = 0 \,. \En{2.7}\EEQ
whereas Eq.~\E{2.3}(b) allows to eliminate one of the components of the electric field,
\BEQ (E_x/v^2)' + {\rm i\hs} k E_z/v^2 = 0 \,,\En{2.8}\EEQ
and Eq.~\E{2.5} gives the expressions for the magnetic field in terms of the electric field 
solutions of the ODEs~\E{2.7},
\BEQ H_x = - (k/\omega) E_y \,, \quad H_y = (k/\omega) E_x + ({\rm i}/\omega) E_z' \,, \quad 
H_z = - ({\rm i}/\omega) E_y' \,.\En{2.9}\EEQ
The equations~\E{2.7}--\E{2.9} determine the solutions of this one-dimensional problem.

An obvious solution is seen at once, viz.\ the \IT{TE mode} with $E_x = E_z = 0$, $H_y = 0$ and 
described by the second order ODE for the transverse (with respect to the propagation 
direction of the waves) component of the electric field,
\BEQ \frac{\ds{d^2}E_y}{\ds d{x^2}}  + (\omega^2/v^2 - k^2) E_y = 0 \,, \En{2.10}\EEQ
or by the  equivalent system of first order ODEs,
\BEQ
\left\{\;\begin{array}{@{}l@{}l}
\frac{\ds d E_y}{\ds dx} - {\rm i} \hs\omega H_z = 0 \,, \\[3mm]
\frac{\ds d H_z}{\ds dx} - ({\rm i}/\omega) (\omega^2/v^2 - k^2) E_y = 0 \,.
\end{array} \right. \En{2.11}\EEQ
whereas $ H_x = - (k/\omega) E_y$ is just an algebraic consequence. The other solution, for 
the \IT{TM mode} with $H_x = H_z = 0$, $E_y = 0$, is described by an analogous second order 
ODE for the transverse component $B_y$ of the magnetic field (or the  equivalent system of 
first order ODEs in terms of $H_y$ and $E_z$), whereas the expression for $E_x$ is just an 
algebraic consequence. These equations are more conveniently obtained from the mirror 
images of Eqs.~\E{2.4}--\E{2.5} where the roles of $\bfE$ and $\bfH$ are exchanged, however 
with different dependencies on $v(x)$. As a result, the spectrum of the TM modes differs 
slightly from the spectrum of the TE modes. We will restrict the analysis to the TE modes, but 
it is obvious that analogous results would be obtained for the TM modes. In conclusion, the 
full electromagnetic spectral wave problem is described by two second order ODEs for two 
slightly different modes, whereas the full MHD spectral wave problem (discussed in 
Sec.~\S{5}) is described by only one second order ODE, but for three completely different 
modes (slow magneto-sonic, Alfv\'en, and fast magneto-sonic) with three different 
phase velocities. Also notice that for real $\omega^2$ the spectral equation~\E{2.10} is of the standard Sturm--Liouville type with monotonicity of the 
eigenvalues with the number of nodes of the internal part of the eigenfunctions, whereas the corresponding MHD equation is much more complicated. Of course, calling Eq.~\E{2.10} a \IT{second order} ODE and 
Eqs.~\E{2.11} \IT{first order} ODEs is only loosely speaking since, for the numerical computation 
of the complex leaky mode spectra, the number of ODEs  actually should be multiplied by two 
for the real and imaginary components of the eigenfunctions involved. Those components will 
be distinguished by the subscripts $1$ and $2$.

\subsection{Boundary value problem for TE modes}{\Sn{2.3}}

Since the TE modes are fully described by the electric field component $E_y$, with the auxiliary magnetic field component $H_z$, it is convenient for the algebra to simply write $E \equiv E_y$ and $H \equiv H_z$ and to distinguish the  outside variables from the inside ones by a hat, $\hat{E}$ and 
$\hat{H}$. With the prescription~\E{2.2} of the phase velocity $v(x)$, the spectral 
equation~\E{2.10} is valid everywhere. Dropping the exponential factor $\exp{\rm i}(k z - 
\omega t)$, the  internal solution is then written as 
\BEQ E = E(x) \,,\quad H = H(x) = - ({\rm i}/\omega) E'(x) \qquad\hbox{($|x| \le 
1$)} \,.\En{2.12}\EEQ 
Since $v(x)$ is assumed to be arbitrary inside the slab, these expressions are to be determined numerically. Outside the slab, $v(x) = 1$, and the external solution is trivial:
\BEQ \hat{E}(x) = \hat{\alpha}\hs{\rm e}^{{\rm i}\hat{q}x} + \hat{\beta}\hs{\rm e}^{-{\rm i} 
\hat{q}x} \,,\quad \hat{H}(x) = (\hat{q}/\omega) (\hat{\alpha}\hs{\rm e}^{{\rm i}\hat{q}x} - \hat{\beta}\hs{\rm e}^{-{\rm i} 
\hat{q}x}) \qquad\hbox{($|x| > 1$)} \,. \En{2.13}\EEQ
If $v(x)$ is assumed to be constant inside the slab, the internal solution also becomes trivial, of the same form as Eq.~\E{2.13} without the hats, and the standard solutions for leaky modes~\R{Huang2023} are obtained. In that case, $\beta = \pm \alpha$ and $\hat{\beta} = 0$, and the jump of the phase velocity at $|x| = 1$ produces the final dispersion equation. However, such an 
assumption is unduly restrictive since problems of genuine interest do not permit this 
simplification. Also, in the corresponding MHD problem this would brush the central Alfv\'en 
singularity under the carpet. Actually, the main problem to be solved here is not the solution 
of differential equations (excellent and accurate numerical techniques exist for this purpose), 
but the satisfaction of the associated  boundary conditions, i.e.\ the solution of the eigenvalue problem. 

For that solution, the complex eigenvalue $\omega\equiv \sigma + {\rm i}\hs\nu$ and the 
real longitudinal wave number $k$ will be treated as known parameters, so that the real and 
imaginary parts of the complex transverse wave number $\hat{q}\equiv \hat{q}_1 + {\rm 
i}\hs\hat{q}_2$  are to be found from the \IT{inverse external dispersion equation}: 
\BEQAR
\oms &=& \hat{q}^2 + k^2 \quad\Rightarrow\quad \hat{q} = \pm \sqrt{\oms - k^2} = \pm A 
\hs{\rm e}^{{\rm i}\hs\varphi} \,,  \En{2.14}\\[2mm]
&&A \equiv (u^2 + v^2)^{1/4}\,,\quad \varphi \equiv \half \arctan(v/u) \,,\qquad u \equiv 
\sigma^2 - \nu^2 - k^2 \,,\quad v \equiv 2 \hs\sigma \nu \,. \non\EEQAR
The central BC, given in Eq.~\E{2.24} below, matching the numerical internal solution~$\{E(x), H(x)\}$ to the explicit external solution~$\{\hat{E}(x), \hat{H}(x)\}$ at $x = 1$, implies matching something like the `internal wave number' $q$ to the external wave number $\hat{q}$. For constant phase velocity, that internal wave number $q$ would follow 
from the internal dispersion equation $\oms = (q^2 + k^2) v^2$. Here, it simply will be an 
implicit property of the internal solution $\{E(x),B(x)\}$ of the ODEs~\E{2.11} which gradually 
transform the wave lengths due to the variation of $v(x)$. Through the definitions~\E{2.14} for 
$A$, $\varphi$, $u$ and~$v$, the expressions for $\hat{q}_1$ and $\hat{q}_2$ will be different 
for the different parts of the complex $\omega$-plane separated by the lines $\sigma = 0$, 
$\nu = 0$, and the hyperbola $\sigma^2 -\nu^2 = k^2$ (see Fig.~\F{2}). In particular, the 
choice of the $\pm$ sign is dictated by the fact that we wish to associate the coefficient 
$\hat{\alpha}$ with outgoing waves and the coefficient $\hat{\beta}$ with ingoing waves. It is 
indicated in the figure for the different areas of the complex $\omega$-plane. 


To exhibit the propagation properties of the waves associated with $\hat{\alpha}$ and $\hat{\beta}$, it is expedient to restore the time dependence and to split real and 
imaginary parts of the exponential factors of the external solutions:
\BEQ \hat{E}(x,t) = \hat{\alpha}\hs{\rm e}^{-\hat{q}_2 x + \nu t}\hs{\rm e}^{\hs{\rm 
i}(\hat{q}_1 x - \sigma t)} + \hat{\beta}\hs{\rm e}^{\hs\hat{q}_2 x + \nu t}\hs{\rm e}^{\hs-
{\rm i}(\hat{q}_1 x + \sigma t)} \,,\En{2.15}\EEQ
and similarly for $\hat{H}(x,t)$. For positive values of 
$\hat{q}_1$ and $\sigma$, the factor $\hat{\alpha}$ then corresponds to 
the outwardly propagating part of the waves, and  the factor $\hat{\beta}$ corresponds to the inwardly propagating part. We may restrict the 
computation of the solutions to $x \ge 0$ since the complete functions, including the behaviour 
for $x < 0$, simply follows from the symmetry imposed at the origin, expressed by the BCs~\E{2.23} below. 
For \IT{leaky modes} propagating exclusively in the $x$-direction ($k = 0$), as in 
\Ron{Huang2023} and our previous paper~\R{GKP2023}, the relation between $\hat{q}$ and 
$\omega$ is obvious: $\hat{q}_1 = \sigma$ and $\hat{q}_2 = \nu$. In that case, the outward 
part (with $\hat{\alpha}$) is exponentially growing in space for $x \rightarrow \infty$ and 
exponentially damped in time for $t \rightarrow \infty$, and for the inward part (with 
$\hat{\beta}$) this behavior occurs for negative values of $x$ and $t$. Assuming an \IT{open 
boundary} at $x = L_x \rightarrow \infty$ which does not reflect waves, it is natural to impose 
the condition $\hat{\beta} = 0$. This implies that the external solution behaves as 
$\hat{\alpha} \exp(-\hat{q}_2 x)$, so that it explodes at infinity for $\hat{q}_2 < 0$, which is 
the characteristic feature of leaky modes. This infinity, the reason for concern  
about the physics of these modes, will be discussed in Sec.~\S{4}.

For \IT{conservative waves}, with real frequencies, propagating in the~$z$-direction ($k \ne 
0$), infinities for $x \rightarrow \infty$ are avoided, even with an open boundary, since $\nu = 
0$ and the values of $\hat{q}$ are then restricted by another condition. In that case, the 
square root sign in the inverse dispersion equation~\E{2.14} gives rise to the following 
distinction of the propagation properties of the external waves:
\BEQ
\hat{q} = \hat{q}_1 + {\rm i}\hat{q}_2 \,,\;\;
\left\{\begin{array}{@{}l@{}l}
\;\hat{q}_1 = 0 \,,\quad\hat{q}_2 = - \sqrt{k^2 - \sigma^2} \qquad\hbox{(for $|\sigma| < k\hs$: 
evanescense)} \,, \\[3mm] 
\;\hat{q}_1 = \sqrt{\sigma^2 - k^2}  \,,\quad\hat{q}_2 = 0 \hor{3}\qquad\hbox{(for $|\sigma| \ge k \hs$: 
propagation)} \,.
\end{array} \right. \En{2.16}\EEQ
Evanescence in the external region ($|x| \ge 1$) does not imply evanescence in the internal 
region:  a discrete spectrum of so-called \IT{bound modes} may be found there, see 
\Ron{SL1983}, p.~206. The first expression is also valid for the \IT{leaky modes} in the $u < 0$ part of the $\omega$-plane, i.e.\ in the so-called \IT{cutoff regions} $1$, $4$ (and $1'$, $4'$) of Fig.~\F{2}. However, the external part of the waves is then explosively growing. It should be noticed also that the description ``propagative'', while appropriate for leaky modes, is actually a misnomer for the conservative modes since these modes are standing waves that do not propagate at all, as we will see shortly. 

To characterise the different cases, and to obtain the pertinent external BCs, we construct the flow of electromagnetic energy, i.e.\ the \IT{Poynting flux}, for $|x| > 1$:
\BEQ \hat{\bfS} \equiv \hat{\bfE} \times \hat{\bfH}^* = \hat{E}(x) \hat{H}^*(x) \hs\bfe_x + (k/\omega^*) \hat{E}(x) \hat{E}^*(x) \bfe_z \,.\En{2.17}\EEQ
Because the system is homogeneous in the~$z$-direction, the contribution of $\hat{S}_z$ just represents a constant flow of energy in that direction, which is of no interest for the present boundary value problem (BVP). Hence, to study the external solutions, the  expression for the Poynting flux may be restricted to the $x$-component:
\BEQ \hat{S}_x(x) = (\hat{q}^*/\omega^*)
\hs\big{[}\hs |\hat{\alpha}|^2
\hs{\rm e}^{- 2\hat{q}_2 x} - |\hat{\beta}|^2
\hs{\rm e}^{2\hat{q}_2 x} - \hat{\alpha} \hat{\beta}^* \hs{\rm e}^{2{\rm i}\hat{q}_1 x} + \hat{\alpha}^* \hat{\beta} \hs{\rm e}^{-2 {\rm i}\hat{q}_1 x} \big{]} \,,\En{2.18}\EEQ
where  the choice of $\hat{\alpha}_{1,2}$ and $\hat{\beta}_{1,2}$ represents different physical problems. This implies\\[2mm]
-- for conservative bound modes (real, $|\sigma| < k$, $\hat{q}_1 = 0$):
\BEQ \hat{\alpha} = 0 \;\;\Rightarrow\;\; {\rm Re}(\hat{S}_x) = 0 \,,\;\; {\rm Im}(\hat{S}_x) = (\hat{q}_2/\sigma)\hs |\hat{\beta}|^2\hs{\rm e}^{2\hat{q}_2 x}|_{x \rightarrow \infty} \rightarrow 0 \,,\En{2.19}\EEQ
-- for conservative `propagating' modes (real, $|\sigma| \ge k$, $\hat{q}_2 = 0$):
\BEQ \hat{\alpha} =  \hat{\beta}\hs{\rm e}^{-2{\hs\rm i}\hs\varphi} \;\;\Rightarrow\;\; {\rm Re}(\hat{S}_x) = 0 \,, \;\; {\rm Im}(\hat{S}_x) = - 2 (\hat{q}_1/\sigma)\hs |\hat{\beta}|^2\hs\sin2(\hat{q}_1 x -\varphi)|_{\rm finite \;all \;x}\,,\En{2.20}\EEQ
-- for cutoff leaky modes (most real, $|\sigma| < k$, but some imaginary, $|\nu| < k$, $\hat{q}_1 = 0$)\hs: 
\BEQ \hat{\beta} = 0 \;\;\Rightarrow
\left\{\begin{array}{@{}l@{}l}
\,\nu = 0: \;{\rm Re}(\hat{S}_x) = 0 \,, \;\; {\rm Im}(\hat{S}_x) = - (\hat{q}_2/\sigma)\hs |\hat{\alpha}|^2\hs{\rm e}^{-2\hat{q}_2 x}|_{x \rightarrow \infty} \rightarrow \infty \,,\\[4mm]
\,\sigma = 0: \;{\rm Re}(\hat{S}_x) = (\hat{q}_2/\nu)\hs |\hat{\alpha}|^2\hs{\rm e}^{-2\hat{q}_2 x}|_{x \rightarrow \infty} \rightarrow \infty \,,\; {\rm Im}(\hat{S}_x) = 0 \,,\end{array} \right. \En{2.21}\EEQ
-- for propagating leaky modes (complex, $\sqrt{\sigma^2 - \nu^2} \ge k$\,, $\hat{q}_1 \ne 0$\,, $\hat{q}_2 \ne 0$):
\BEQ \hat{\beta} = 0 \;\;\Rightarrow\;\; \hat{S}_x = (\hat{q}^*/\omega^*)
\hs|\hat{\alpha}|^2\hs{\rm e}^{- 2\hat{q}_2 x} \;\Rightarrow\,\; {\rm Re}(\hat{S}_x)|_{x \rightarrow \infty} \;\&\; {\rm Im}(\hat{S}_x)|_{x \rightarrow \infty} \rightarrow \infty \,.\En{2.22}\EEQ
Consequently, the two kinds of modes  may be distinguished  by whether both components of~$\hat{S}_x$ are finite (conservative modes) or at least one of them is unbounded (leaky modes). We will discuss in Sec.~\S{4} when and why the latter modes are still acceptable in a number of physical applications. For all cases with real eigenvalues,  ${\rm Re}(\hat{S}_x) = 0$  for all $|x| > 1$, which represents the rate of energy change  averaged over an oscillation period. Apparently, this condition does not distinguish conservative from leaky modes since  a sub-spectrum of real discrete leaky modes also satisfies it according to Eq.~\E{2.21}. We will come back to this when discussing Fig.~\F{4} in Sec.~\S{3.3}. 

Summarizing: the  boundary value problems (BVPs) for the numerical solution of the spectral equation~\E{2.10} imply imposing the following boundary conditions:
\BEQAR
\hor{-10}&&\hbox{(a)~at $|x| = 0$:} \quad E(0) = 0 \quad\hbox{\IT{(odd modes)}}\,, \quad 
\hbox{or}\quad H(0) = 0 \quad\hbox{\IT{(even modes)}}  \,,\En{2.23}\\[2mm]
\hor{-10}&&\hbox{(b)~at $|x| = 1$:} \quad E(1) = \hat{E}(1) \;\; \hbox{and}\;\; H(1) = 
\hat{H}(1) \,,\En{2.24}\\[2mm] 
\hor{-10}&&\hbox{(c)~at $|x| \rightarrow \infty$\,,}\; 
\left\{\begin{array}{@{}l@{}l}
\,(c1)\; \hbox{\IT{CNMs}:}  \;\;\hat{\alpha} = 0 \;\;\hbox{(for $\sigma^2 - \nu^2 < k^2$)}\,,\quad |\hat{\alpha}| = |\hat{\beta}| \;\;\hbox{(for $\sigma^2 - \nu^2 \ge k^2$)}\,,\\[4mm]
\,(c2)\; \hbox{\IT{QNMs}:} \;\;\hat{\beta} = 0\,.
\end{array} \right. \En{2.25}\EEQAR
At this point, we have introduced our main focus with the split into the two distinct BVPs 
indicated by the labels (c1) for \IT{conservative normal modes} (CNMs), with real eigenvalues, 
and (c2) for \IT{leaky, quasi-normal modes} (QNMs), with real or complex eigenvalues. We exploit the acronym QNMs for the leaky modes following standard terminology, see e.g.\  \Ron{Colom2018} and  \Ron{Sauvan2022}. The construction of the solutions of these BVPs involves matching the numerical internal solution~$\{E(x), H(x)\}$, determined by the BCs~\E{2.23}, to the explicit external solution~$\{\hat{E}(x), \hat{H}(x)\}$, determined by the BCs~\E{2.25}, and subsequently trying to satisfy the two BCs~\E{2.24}. Satisfaction of the first one (on the electric field) can always be obtained by adapting the amplitude of the external part of the solution, but the second one (on the magnetic field) can only be satisfied for certain values of~$\omega$, viz.\ \IT{the eigenvalues.} Solution of this eigenvalue problem will be accomplished by means of the method of the Spectral Web, introduced in the following sub-section. 

Finally, it is to be noted that, by virtue of the explicit expressions~\E{2.13} for the external solution, satisfaction of the BCs~\E{2.25} for $|x| \rightarrow \infty$ directly transfers to explicit expressions for the variables $\hat{E}(1)$ and $\hat{H}(1)$ that are exploited in the Spectral Web method. Also, note that apostrophes have been put on `propagating' for the real conservative modes in the region $|\sigma| \ge k$, with the property~\E{2.20}. That property guarantees perfect balance between the outgoing and ingoing waves so that these modes do not propagate at all!  This implies that the appropriate terminology is \IT{standing conservative modes}, distinguishing them from \IT{propagating leaky modes}. This distinction is crucial for the different normalisations discussed in Sec.~\S{4}.

\subsection{Spectral Web for electrodynamic waves in dielectric media}{\Sn{2.4}}

Our method to construct the eigenvalue spectra is termed the \IT{Spectral Web}, analogous to the Spectral Web for MHD plasmas introduced by~\Ron{Goed2018} and effectively applied 
to accretion disks about black holes~\R{GK2022} discovering the new Super-Alfv\'enic Rotational 
Instabilities (SARIs) described in that paper. Its counterpart is here developed for electrodynamic waves in dielectric media.

 Satisfaction of the central BCs~\E{2.24} for arbitrary complex values of $\omega$ implies that the four coefficients $\{\hat{\alpha}_{1,2}, \hat{\beta}_{1,2}\}$ of the external solution~\E{2.13} are completely determined by the four boundary values $\{E_{1,2}(1), H_{1,2}(1)\}$ of the internal solutions~\E{2.12}: 
\BEQ \hat{\alpha} = \half {\rm e}^{-{\rm i}\hat{q}} \big[E(1) + (\omega/\hat{q}) H(1)\big] \,,\quad
\hat{\beta} = \half {\rm e}^{{\rm i}\hat{q}} \big[E(1) - (\omega/\hat{q}) H(1)\big]\,. \En{2.26}\EEQ
However, for arbitrary values of $\omega$, this conflicts with also satisfying the outer BCs~\E{2.25}, since they involve a restriction on the  coefficients $\{\hat{\alpha}_{1,2}, \hat{\beta}_{1,2}\}$.  This implies that, eventually, the values of~$\omega$ have to be restricted to the subset of complex frequencies which constitute the spectrum of eigenvalues. We temporarily set this conflict aside and just compute, for arbitrary complex values of $\omega$, the mismatch of satisfying both BCs~\E{2.24}. As already mentioned, the jump of the electric field can always be imposed to vanish, satisfying the first part of the BCs~\E{2.24}, 
\BEQ
\doublel{E(1)}\doubler \equiv \hat{E}(1) - E(1) = 0 \quad\Rightarrow\quad \hat{E}(1) 
\equiv \hat{\alpha}\hs{\rm e}^{{\rm i}\hat{q}} + \hat{\beta}\hs{\rm e}^{-{\rm i} \hat{q}} = 
E(1) \,,  \En{2.27} \EEQ
but the jump of the magnetic field does not vanish under the same conditions,
\BEQ
\doublel{H(1)}\doubler \equiv \hat{H}(1) - H(1)\ \equiv (\hat{q}/\omega) 
(\hat{\alpha}\hs{\rm e}^{{\rm i}\hat{q}} - \hat{\beta}\hs{\rm e}^{-{\rm i} \hat{q}}) - H(1) \ne 
0 \,. \En{2.28}\EEQ
Consequently, iterative determination of the eigenvalues requires a numerical procedure  producing the required limit $\doublel{H(1)}\doubler \rightarrow 0\hs$. As we will demonstrate, construction of the 
Spectral Web is  extremely well suited for this purpose.

This method  involves  computation of the \IT{complementary Poynting flux} across 
the surface of discontinuity. It is defined as the flux of electromagnetic energy per 
unit area, averaged over the ignorable coordinates $y$ and $z$, that should be injected or extracted at that surface to produce the pertinent
discontinuity of the electromagnetic variable: 
\BEQ S_{\rm com} \equiv \frac{1}{L_y L_z} \int_{-L_y}^{L_y} \int_{-L_z}^{L_z} 
\doublel{S_x}\doubler {\hs}dy{\hs}dz = \doublel{S_x(1)}\doubler 
= E(1) \hs\doublel{H^*(1)}\doubler \,.\En{2.29}\EEQ
Clearly, this complex quantity precisely represents the partial mismatch of the two expressions~\E{2.27} and \E{2.28}. It should vanish to satisfy the full BVP. The method consists of computing $S_{\rm com}$ for a large range of values of $\omega$ in a 
relevant region of the complex $\omega$-plane, and determining the paths on which the 
real and imaginary parts of $S_{\rm com}$ vanish separately. Eigenvalues are found at the 
intersections of the two paths, where  both components of $S_{\rm com}$ vanish. Obviously, both BCs~\E{2.24} will then be satisfied. Indicating the real and imaginary parts of E(1) by $E_1$ and $E_2$, and similarly for $B(1)$, the explicit expressions for the two paths of the Spectral Web read:
\BEQ \left. \begin{array}{l}
S_{{\rm com},1} \equiv {\rm Re}(S_{\rm com}) = E_1 \hs\doublel{H_1}\doubler + E_2 \hs\doublel{H_2}\doubler = 0 \quad\hbox\IT{(solution path)}\\[4mm] 
S_{{\rm com},2} \equiv {\rm Im}(S_{\rm com}) = E_2 \hs\doublel{H_1}\doubler - E_1 \hs\doublel{H_2}\doubler = 0 \quad\hbox\IT{(conjugate path)} 
\end{array} \right\} \;\;\Rightarrow\; S_{\rm com}(\omega) = 0 \,. \En{2.30}\EEQ
Here, the numerical values of $E_{1,2}$ and $H_{1,2}$ and the analytical values of $\hat{H}_{1,2}$, for the particular choice of the parameters $\hat{\alpha}$ and $\hat{\beta}$ dictated by Eqs.~\E{2.25} of the BVP involved, are to be substituted.

Each path also has a physical meaning by itself. Considering an external source exciting the waves at $|x | = 1$, for the solutions on the solution path the energy flows back and forth between the source and the system without changing the total energy averaged over an oscillation period (the coupled system is conservative), whereas that energy does change for the solutions on the conjugate path (the coupled system is non-conservative). For the eigenvalues at the intersections of the two paths, system and source are uncoupled: no complementary Poynting flux is needed, $S_{\rm com} = 0$. However, notice that spurious solutions with $S_{\rm com} = 0$ also occur, viz.\ when accidentally $E(1) = 0$ for some frequency. Those `solutions' are easily identified and eliminated by demanding that $|E(1)| \ne 0$. In the  Spectral Webs shown the next section, the solution paths are coloured red and the conjugate paths blue. 

The attraction of the Spectral Web method is that it provides 
a rough graphical impression of the structure of the spectrum already by means of a coarse 
grid in the $\omega$-plane, giving an indication of where cluster points and continuous 
spectra are to be expected. It may be refined to any degree by just zooming in onto 
a relevant part of the $\omega$-plane, and then iterating along either one of the paths, e.g.\ the solution path, to obtain the eigenvalues to the accuracy desired. That iteration is  similar to the standard shooting method for the Sturm--Liouville problem, where the complex eigenvalue variation is now also one-dimensional since confined to the solution path.

\ver{10}

\section{Spectra of conservative and leaky TE modes}{\Sn{3}}

\subsection{Preliminaries}{\Sn{3.1}}

To compute the eigenvalue spectra of the modes,  the phase velocity~$v(x)$ 
in the inner region should be specified. We choose the following two-parameter family of functions:
\BEQ v(x) = 
\begin{cases}
\,\delta \hor{62}\qquad\hbox{($0 \le |x| \le 1 - \epsilon$)}\,, \\[2mm]
\,\half(1 + \delta) + \half (1 - \delta) \sin\hs[(|x| - 1 + \half\epsilon)\pi/\epsilon\hs] 
\qquad\hbox{($1 - \epsilon \le |x| \le 1$)} \,.
\end{cases}\En{3.1} \EEQ
Here, $\delta$ represents the phase velocity $v(x) = \delta < 1$ in the innermost region of 
the slab and $\epsilon$ is the size of the outermost region of the slab where $v(x)$ smoothly 
transforms to the value $1$ (i.e.\ the velocity of light in vacuum). For $\epsilon = 0$, the 
standard discontinuous leaky mode model~\R{Huang2023} is obtained. In the present calculation, the parameters $\delta$ and $\epsilon$ may  have any value between $0$ and $1$. Of course, the above choice of the function $v(x)$ is just for the sake of definiteness, any other choice of a smooth function transforming from $\delta$ to $1$ will be acceptable.

Actually, that assumption of smoothness is only made for the sake of the comparison with the MHD modes in the solar corona, where discontinuous backgrounds are always crude approximations of gradual variations of background variables, as discussed in Secs.~\S{6} and following. For dielectric materials consisting of several layers with different values of the dielectric constant, even frequency-dependent ones, the internal numerical solution can be obtained with the same method as discussed above by joining the different layers with internal BCs imposing continuity of both $E$ and $H$. It is only in the final construction of the eigenvalues that the discontinuity of $\doublel H \doubler$ at $|x| = 1$ is eliminated by the iterative procedure of the Spectral Web.

In the following two sub-sections,  the Spectral Web solution method will be applied to the conservative and leaky modes corresponding to  the two outer BCs~\E{2.25} indicated by (c1) and (c2). It is to be noted that the solution~\E{2.14} of the inverse external dispersion equation provides the explicit values of the parameters $\hat{q}_1$ and $\hat{q}_2$ of the external solutions~$\hat{E}(x)$ and $\hat{H}(x)$. Hence, in this case, any integration over the external interval~$|x| > 1$ can be carried out by hand. Numerical integration of the ODEs~\E{2.11} over the interval~$|x| \le 1$, and subsequent substitution of the  values of $E(1)$ and $H(1)$ into the expression~\E{2.29} for the complementary Poynting flux, provides the basic step for the solution of the  eigenvalue problem. The rest is just a graphical construction of the  two paths in the complex $\omega$-plane, or the more precise determination of the eigenvalues by iterating along one of the paths of the Spectral Web. 

\subsection{Conservative normal modes}{\Sn{3.2}}

The spectra of  conservative normal modes are shown in Figs.~\F{3}(a) for the even variety, and in Fig.~\F{3}(b) for the odd variety. For all calculations of the present section, the parameters $k = 4.0$, $\delta = 0.25$ and $\epsilon = 0.02$ were chosen for the purpose of illustration, and also for comparison with earlier calculations by \Ron{Armitage2014}. Collecting the expressions for the \IT{conservative} ($\nu = 0$) \IT{bound modes} ($|\sigma| < k$), i.e.\ in the cutoff region, yields
\BEQ \hat{\alpha} = 0\,,\;\; \hat{q} = - {\rm i}\sqrt{k^2 - \sigma^2} \quad\Rightarrow\;\; S_{\rm com} = E(1)\big[\hs{\rm i}\hs(\hat{q}_2/\sigma)\hs E^*(1) - H^*(1)\big]\,,\En{3.2}\EEQ
where the contribution $S_x(1) \equiv E(1) H^*(1)$ of the internal solution is also imaginary since the basic ODE~\E{2.10} has only  real coefficients. This implies that one could choose initial data such that $E(x)$ is real and $H(x)$ is imaginary. Choosing complex initial data would not change this conclusion. Hence, $S_{\rm com,1} = 0\hs$: the solution path is along the real axis, whereas the iterative procedure should produce $S_{\rm com,2} \rightarrow 0\hs$. For this procedure, there is actually no need to construct the Spectral Web paths in the complex $\omega$-plane, since the solutions  can be obtained by a standard shooting method.  The resulting discrete spectra are shown in Fig.~\F{3} for the mentioned values of $\delta$ and $\epsilon$.  In the positive frequency range $0 < \sigma < k$, they consist of five even and five odd bound modes, and an equal number of even and odd modes in the negative frequency range. The green lines in Figs.~\F{3} and \F{4} indicate the boundaries between the cutoff and propagating regions given by the hyperbola $\sigma^2 - \nu^2 = k^2$ of Fig.~\F{2} (nearly straight lines because of the small vertical section of the $\omega$-plane).

As is well known, the \IT{conservative} ($\nu = 0$) \IT{`propagating'}, or rather, \IT{standing waves} ($|\sigma| \ge k$)  constitute \IT{continuous spectra} of all real frequencies in those two ranges (see Fig.~\F{3}). Hence, the Spectral Web machinery is not needed either to construct these spectra. This should imply that for all these frequencies there is no contradiction in satisfying the full BVP, guaranteed by the expressions~\E{2.26}, and restricting $\hat{\alpha}$ and $\hat{\beta}$ by the condition~\E{2.20}. This may be proved by choosing initial data such that $E(1) = E_1$, $H(1) = {\rm i\hs} H_2$, so that the expressions~\E{2.26} become
\BEQAR
 &&\hat{\alpha} = \half {\rm e}^{-{\rm i}\hat{q}_1} \big[E_1 + {\rm i\hs}(\sigma/\hat{q}_1) H_2\big] \,,\quad
\hat{\beta} = \half {\rm e}^{{\rm i}\hat{q}_1} \big[E_1 - {\rm i\hs}(\sigma/\hat{q}_1) H_2\big]\,,\;\;\hbox{with}\;\; \hat{q}_1 = \sqrt{\sigma^2 - k^2}\,, \non\\[2mm]
&&\Rightarrow\;\; \hat{\alpha} =  \hat{\beta}\hs{\rm e}^{-2{\hs\rm i}\hs\varphi}, \;\:
|\hat{\alpha}| = |\hat{\beta}| = \half \sqrt{E_1^2 + (\sigma H_2/\hat{q}_1)^2} \,,\;\;
\varphi = \arctan\big[\sigma H_2/(\hat{q}_1 E_1)\big]\,,\En{3.3}\EEQAR
\IT{QED.} Again, this derivation will not change by choosing complex initial data instead. In conclusion, all frequencies satisfying $|\sigma| \ge k$ are admitted. The `improper eigenfunctions' associated with these modes are non-square-integrable because of the infinite domain (as in quantum mechanics of free particles). Notice that, according to Eq.~\E{2.20}, no energy is lost at the ends, for $x \rightarrow \infty$, even though the system is open there.


\subsection{Quasi-normal, leaky modes}{\Sn{3.3}}

The spectra of leaky modes are shown in Figs.~\F{4}(a) for the even and in Fig.~\F{4}(b) for the odd variety. The parameters are the same as for the conservative modes in Sec.~\S{3.2}.
According to the expressions~\E{2.21}, \IT{cutoff leaky modes} ($\sigma^2 - \nu^2 \le k^2$) are 
restricted to lie either on the real $\sigma$-axis or on the negative part of the $\nu$-axis:
\BEQ \hat{\beta} = 0\,,\;\; \hat{q} = - {\rm i}\sqrt{k^2 - \omega^2} \quad\Rightarrow\;\; S_{\rm com} = E(1)\big[f E^*(1) - H^*(1)\big]\,,\;\; f \equiv \left\{\begin{array}{@{}l@{}l}
 -{\rm i}\hs\hat{q}_2/\sigma \;\hbox{($\nu = 0$)}\\[1mm]
\hor{4}\hat{q}_2/\nu \;\hbox{($\sigma = 0$)} \end{array} \right..\En{3.4}\EEQ
Hence, they may be calculated by shooting, as for the conservative bound modes, since the basic 
ODE~\E{2.10} depends only on the squared frequency $\omega^2$ ($= \sigma^2$ or $-\nu^2$), which is real. (In Figs.~\F{4}, the Spectral Web contours in the cutoff region are shown anyway to exhibit their peculiar continuation in the propagating region.) Most of the cutoff leaky modes lie on the real $\sigma$-axis where $S_{\rm com}$ is imaginary, but some lie on the imaginary $\nu$-axis where $S_{\rm com}$ is real. In this case there is only one of the latter modes, viz.\ the degenerate $n = 0$ even mode for $\nu \approx -0.04$. The  $n = 0$ odd modes lie too close to the origin to see that they are actually non-degenerate: $\sigma \approx \pm 0.04$. (Note that the scale along the $\sigma$-axis is much larger than along the $\nu$-axis.) The near vertical blue curves of the conjugate paths actually form closed loops faraway in the $\omega$-plane such that each loop contains just one eigenvalue and one false solution on the real axis. This guarantees monotonicity with arc length of the pertinent component of $S_{\rm com}$ in between those points, as generally proved for the analogous quantity $W_{\rm com}$ for MHD waves~\R{Goed2018}.

Finally, \IT{propagating leaky modes} ($\sigma^2 - \nu^2 > k^2$) have complex values of $\omega$, and hence of $\hat{q}$, so that the full Spectral Web machinery is needed to calculate their eigenvalues and eigenfunctions. From Eq.~\E{2.22} we  get the following expression for the complementary Poynting flux exploited in this calculation:
\BEQ \hat{\beta} = 0\,,\;\; \hat{q} = \pm \sqrt{\oms - k^2} \quad\Rightarrow\;\; S_{\rm com} = E(1)\big[\hs(\hat{q}^*/\omega^*)\hs E^*(1) - H^*(1)\big]\,.\En{3.5}\EEQ
The plots of the resulting complex eigenvalues are shown in Fig.~\F{4}(a) for the even modes and in Fig.~\F{4}(b) for the odd modes. Whereas strictly speaking,  the usual labeling by node counting of the eigenfunctions fails for complex eigenvalue problems, the numbering of the complex modes ($n \ge 6$) can be continued here from the real ones ($n = 1,  \ldots 5$) by means of the  mentioned monotonicity properties along the paths of the Spectral Web. Nevertheless, it appears to be just a  continuation of the numbering for the real ones. 

\subsection{Different spectral methods}{\Sn{3.4}}

The  spectra shown in Figs.~\F{3} and \F{4} agree, qualitatively, with the spectra shown in Fig.~1 of \Ron{Armitage2014}. This reference assumes constant phase velocity in the dielectric slab (of course with a jump at the boundary) as a first approximation to more complicated structures. This paper is one in a sequence of papers~\R{Doost2012,Doost2013,Doost2014} on the so-called {\em Resonant State Expansion} of ever more complicated systems, where the central dielectric slab is modified, e.g.\ with layers of different values of the phase velocity. This includes frequency dependent ones~\R{Muljarov-Langbein2016} and even anisotropic magnetic materials~\R{Muljarov-Weiss2018}. Likewise,~\Ron{Upendar2018} investigate the properties of different optical fiber systems with such a method. The resulting complicated spectrum is  computed by means of a perturbation method based on the simple unperturbed constant phase velocity model, exploiting  Greens function techniques (to be discussed in Sec.~\S{4.1} on the initial value problem). This computation requires solving a nonlinear eigenvalue problem with a large number of modes of the unperturbed system. 

In the Spectral Web method, this part of the mentioned procedure is replaced by the numerical integration of the ODEs in the internal region since it automatically adjusts the integration step to enforce continuity of $E(x)$ and $H(x)$ at the boundaries of the different internal domains where a discontinuous phase velocity profile is prescribed. Solution of the resulting spectral problem is then concentrated in solely computing the complementary Poynting flux, and the associated paths in the complex plane, at the final outer boundary $|x| = 1$ of the inhomogeneous region. In the present case, obtaining the external solutions $\hat{E}(x)$ and $\hat{H}(x)$ is completely explicit through the inversion of the external dispersion equation~\E{2.14}. For non-trivial more-dimensional external structures, numerical integration over the external region satisfying the outflow BC would be necessary. This may be performed, for arbitrarily chosen complex values of~$\omega$, virtually without any approximation since the required accuracy of modern numerical solvers may be prescribed. The eigenvalues themselves are then obtained from the zeros of the complementary Poynting flux in the next step of the procedure. 

\subsection{CNM and QNM eigenfunctions}{\Sn{3.5}} 

Characteristic eigenfunctions in, respectively, the cutoff and standing-wave/propagating regions are shown in Figs.~\F{5}(a) and (b) for the conservative modes (labelled by the mode number $n$ in the spectra of Fig.~\F{3}), and in Figs.~\F{6}(a) and (b) for the leaky modes (labelled by the mode number $n$ in the spectra of Fig.~\F{4}). To save space, only the electric fields are exhibited, the associated magnetic fields have been omitted. Their amplitudes are indicated though. Peculiarly, all amplitude changes of the magnetic field in the very narrow region of inhomogeneity (better visible in the wider regions of their MHD counterparts $\Pi$ and $\xi$ in Figs.~\F{9} and \F{10}) are opposite to those of the electric fields.

The conservative modes are divided in bound modes for $|\sigma| < k$, which are oscillatory for $|x| < 1$ but evanescent for $|x| \ge 1$ (Fig.~\F{5}(a)), and standing waves for $|\sigma| \ge k$, which are oscillatory in both regions but with different wave lengths (Fig.~\F{5}(b)). The bound modes are discrete, but the standing waves form continuous spectra since their domain is infinite in $x$. This infinite domain is representative of an open boundary but, since the amplitudes of the standing waves are constant, no energy is lost at infinity!

The leaky modes are also distinguished in two kinds of modes: so-called cutoff modes for $|\sigma| < k$, which are partly oscillatory in the region $|x| < 1$ and exponentially growing for $|x| \ge 1$ (Fig.~\F{6}(a)), and propagating waves for $|\sigma| \ge k$, which oscillate in both regions, again with different wave lengths, but their amplitude is exponentially growing in the external region (Fig.~\F{6}(b)). The explosive growth of the leaky modes for $x \rightarrow \infty$ results in exponentially infinite energy loss there! Associated with this spatial growth is the temporal damping $\nu < 0$ of the modes. Clearly, in contrast to unstable systems where the perturbations are assumed to grow from tiny fluctuations of the background, leaky modes can not appear in this way. They require finite initial perturbations, as discussed in the next section. How the infinite energy flux for $|x| \rightarrow \infty$ is sidestepped for TE modes is the subject of Sec.~\S{4.2}. Whether this `magic' also works for MHD waves in coronal flux tubes will be investigated in Sec.~\S{7}.

Since the exponentially explosive part of the cutoff leaky modes solution is restricted to the external part 
($|x| > 1$), the oscillations of the internal part may be exploited to label the solutions with their number of nodes (never mind that they also may have a large, but not infinite, amplitude for $|x| \le 1$). Significantly, the bound conservative modes are solutions of the same Sturm--Liouvile type ODE~\E{2.10}! This permits to arrange the sequences of discrete eigenvalues of the leaky modes (subscript ${\rm L}$) in an interweaving pattern with those of the conservative bound modes (subscript ${\rm C}$), distinguishing between even and odd (subscripts ${\rm e}$ and ${\rm o}$), according to the following scheme: 
\BEQ |\nu_{{\rm L}{1{\rm e}}}| < |\sigma_{{\rm C}{1{\rm e}}}| < |\sigma_{{\rm L}{1{\rm 
o}}}| < |\sigma_{{\rm C}{1{\rm o}}}| < |\sigma_{{\rm L}{2{\rm e}}}| < |\sigma_{{\rm 
C}{2{\rm e}}}| < |\sigma_{{\rm L}{2{\rm o}}}| < |\sigma_{{\rm C}{2{\rm o}}}| \,\ldots \,(< 
k) \,.\En{3.6}\EEQ
The  explicit eigenvalues of these sequences are given in Table~\T{1}(a). They are completed with the complex eigenvalues of the propagating leaky modes in Table~\T{1}(b). Since the conjugate paths are nearly vertical close to the crossings with the solution path, extremely fast and accurate convergence to those eigenvalues is obtained by just iterating along the relevant conjugate path. The numbers obtained are truncated to five decimal places (more than enough for the present purpose). Of course, the real counterparts of the standing conservative modes are missing since they occupy all real numbers $\ge k$.

The frequency scheme of Table~\T{1} consists of numbers that depend on the specific inhomogeneity {\em of the internal region only:} perfect for classical (non-quantum mechanical) spectroscopy to determine the structure of dielectric materials with external excitation. That is, {\em if only the infinities can be avoided,} which is the subject of the next section.


\section{Excitation of conservative and leaky electromagnetic waves}{\Sn{4}}

\subsection{The initial value problem}{\Sn{4.1}}

So far, we have presented conservative and leaky modes on an equal footing but, obviously, one physical system cannot be described by two different eigenvalue spectra. The pertinent question to be answered is: when is the conservative picture appropriate and when does the leaky mode picture apply? Broadly speaking, the regular conservative picture provides the final, stationary state, response of the system to excitation, whereas the leaky mode picture provides a description of the response of the system in the initial phase when, so to speak, the outward waves do not yet ``know'' that they are on the way to infinity and assumed to grow without bound. The latter description has been accepted for electromagnetic waves for a long time, and for the obvious reason that it is observed, but only rather recently it has been provided with a theoretical description on a par with the standard description of conservative modes in terms of self adjoint operators in Hilbert space; see the tutorial by~\Ron{Sauvan2022}. This distinct description of leaky modes, called ``regularisation'', is the subject of the following sub-section. Before discussing that, we first present what is common to the two descriptions.

To describe the response to excitation of TE modes, the usual formulation of the solution of the initial value problem in terms of Green's functions is most effective. Exploiting the formulation of our textbook, \Ron{GKP2019}, we transform the prospective time-dependent solution $E(x;t)$ to the frequency-dependent variable $\widetilde{E}(x;\omega)$ by means of the \IT{forward Laplace transformation},
\BEQ \widetilde{E}(x;\omega) \equiv \int_0^\infty {\!\!}E(x;t) \hs{\rm e}^{\hs{\rm i}\omega t} d{\hs}t \,.\En{4.1}\EEQ
The original time-evolution equation~\E{2.4}, simplified for the present TE modes to
\BEQ \frac{1}{v^2}\deldeltt{E} - \deldelxx{E} + k^2 E = 0 \,, \En{4.2}\EEQ
is also subjected to that transformation:
\BEQ \ddxx{\widetilde{E}} + (\omega^2/v^2 - k^2) \widetilde{E} = \big({\rm i}\hs\omega E_{\rm i}(x) - \dot{E}_{\rm i}(x)\big)/v^2 \equiv X(x;\omega) \,. \En{4.3}\EEQ
Here, the excitation $X = X(x;\omega)$ is assumed to be known from  prescribed initial data $E_{\rm i}(x) \equiv E(x; t\!=\!0)$ and $\dot{E}_{\rm i}(x) \equiv {\partial E/\partial t}(x; t\!=\!0)$. Those are obtained from the Laplace transform of the first term of Eq.~\E{4.2} by integrating by parts and neglecting the terms multiplying $\exp({\hs{\rm i}\omega t})$ for $t \rightarrow \infty$ since they vanish \IT{in the upper part of the complex~$\omega$-plane}. This restriction will be removed later by means of analytic continuation.

The inhomogeneous ODE~\E{4.3} is solved by the integral
\BEQ \widetilde{E}(x;\omega) = \int_0^\infty G(x,x';\omega) X(x';\omega) d{\hs}x' \,.\En{4.4}\EEQ
where the Green's function $G(x,x';\omega)$ satisfies another inhomogeneous ODE,
\BEQ \Big[ \ddxx{}  + \omega^2/v^2 - k^2 \Big] G(x,x';\omega) = \delta(x - x') \,, \En{4.5}\EEQ
subject to the left BC at $x = 0$, the right BC for $x \rightarrow \infty$,  continuity, $\doublel G \doubler
= 0$, and unit jump of the derivative, $\doublel dG/dx \doubler = 1$, at the joining point $x = x'$. The explicit solution of Eq.~\E{4.5} is then given by the quotient
\BEQ G(x,x';\omega) = \frac{\Gamma(x,x';\omega)}{\Delta(\omega)} \,,\En{4.6}\EEQ
where the numerator and denominator are composed of the left and right solutions,  $U_1(x;\omega)$ and $U_2(x;\omega)$, of the original homogeneous ODE~\E{2.10},
\BEQAR
\Gamma(x,x';\omega) &=& U_1(x;\omega) U_2(x';\omega) H(x' -  x) + U_1(x';\omega) U_2(x;\omega) H(x - x') \,, \nonumber\\[2mm]
\Delta(\omega) &=& U_1(x;\omega) U'_2(x;\omega) - U'_1(x;\omega) U_2(x;\omega) \,,
\En{4.7}\EEQAR
and $H$ here indicates the Heaviside function. The Wronskian $\Delta$ is independent of $x$ as a direct consequence of the differential equation. In this manner, the numerical solution~\E{2.12} and the analytical solution~\E{2.13} of Sec.~\S{2.3} return: $U_1(x;\omega) \equiv E(x;\omega)$ for $|x| \le 1$ and $U_2(x;\omega) \equiv \hat{E}(x;\omega)$ for $|x| > 1$. However, the important difference is that the left solution $U_1$, satisfying the BC~\E{2.23} at $x = 0$, is continuously extended into the right part, and the right solution, satisfying either one of the BCs~\E{2.25} at $x \rightarrow \infty$, is continuously extended into the left part. These extensions demand that the corresponding electric and magnetic fields do not jump at $|x| = 1$, i.e.\ $\doublel U_i \doubler = 0$ and $\doublel U'_i \doubler = 0$, where $i = 1, 2$. In general, this implies that the two solutions do not satisfy the required BCs at the other end, so that $U_1$ and $U_2$ are not eigenfunctions. 

At this point, it is expedient to convert the expression~\E{4.7} for $\Delta$, which may also be called \IT{the dispersion function}, into the physical variables  $E$ and $H$ at the position $x = 1$ (i.e. at the position where the extensions of $U_1$ and $U_2$ are not relevant):
\BEQ\Delta = ({\rm i}/\omega) [E(1;\omega) \hat{H}(1;\omega) - H(1;\omega) \hat{E}(1;\omega)] = ({\rm i}/\omega)E(1;\omega) \doublel H(1;\omega) \doubler \,, \En{4.8}\EEQ
where the arbitrary amplitudes of the two functions $U_1$ and $U_2$ have been chosen to correspond to the same electric field at $x = 1$. For eigenvalues $\omega = \omega_n$ the two functions $U_1$ and $U_2$ become identical (and eigenfunctions), so that 
\BEQ U_1(x;\omega_n) = U_2(x;\omega_n) \equiv E(x;\omega_n) \;\;\;\hbox{for all $x$} \quad\Rightarrow\quad \Delta(\omega_n) = 0 \,,\En{4.9}\EEQ
where, from now on, the physical variable $E(x;\omega_n)$ indicates both the internal numerical part and (dropping the hat) also the external analytical part of the solution. 

As a side remark, note that there is an obvious relation with the complementary Poynting flux~\E{2.29} of Sec.~\S{2}. Since $\doublel H(1;\omega) \doubler = \exp(2\hs{\rm i}\hs\zeta) \doublel H^*(1;\omega) \doubler$, with the  phase angle $\zeta = \zeta(\omega)$, the two are directly related:
\BEQ \Delta(\omega) = f(\omega) S_{\rm com}(\omega)\,,\quad f(\omega) \equiv ({\rm i}/\omega) {\rm e}^{2\hs{\rm i} \zeta(\omega)} \,.\En{4.10}\EEQ
Hence, one may construct Spectral Webs just as well from $\Delta(\omega)$ as from $S_{\rm com}(\omega)$. They will give different solution and conjugate paths in the complex $\omega$-plane but their crossings will occur at the same eigenvalue positions. 

At the eigenvalue frequencies $\omega = \omega_n$, the numerator of the Green's function quotient simplifies to 
\BEQ \Gamma(x,x';\omega_n) = E(x;\omega_n) E(x';\omega_n) \,,\En{4.11}\EEQ
whereas the denominator may be expanded:
\BEQ \Delta(\omega) \approx (\omega - \omega_n) \Big[\frac{\partial\Delta}{\partial\omega}\Big]_{\omega_n} \!\!\!\approx (\omega - \omega_n) A_n\,, \qquad A_n \equiv  \Big[ ({{\rm i}}/{\omega}) E(1;\omega) \hs\frac{\partial\doublel{H(1;\omega)}\doubler}{\partial\omega} \Big]_{\omega_n} \,. \En{4.12}\EEQ
The constants $A_n$ may be computed and then eliminated by normalising the solutions $E_n = E(x;\omega_n)$ such that these constants become equal to $1$, so that they disappear from the Green's function in the neighbourhood of the eigenvalue $\omega_n$:
\BEQ G(x,x';\omega) \approx \frac{E(x;\omega_n) E(x';\omega_n)}{\omega - \omega_n} \,,\En{4.13}\EEQ
The Green's function is now sufficiently explicit in terms of the eigenfunctions to construct the final solution of the initial value problem. 

That final solution involves the \IT{inverse Laplace transformation,}
\BEQ E(x;t) \equiv \frac{1}{2\pi}\int_{C_0} \!\!d{\hs}\omega \hs\hs{\rm e}^{-\hs{\rm i}\hs\omega t} \!\int_0^\infty G(x,x';\omega) {\hs}X(x';\omega) {\hs}d{\hs}x' \,,\En{4.14}\EEQ
where $C_0$ indicates a line from $\hs\omega = -\infty + {\rm i}\hs\nu_0\hs$ to $\hs\omega = \infty + {\rm i}\hs\nu_0\hs$, with $\nu_0 > 0\hs$, in the upper half of the $\omega$-plane. To exploit analytic continuation, this line is connected with a downward/upward contour into the bottom half of the $\omega$-plane, e.g. a semi-circle of radius $R \rightarrow \infty$. The integral over that additional piece vanishes, so that the line integral \E{4.14} over $C_0$ is transformed into a contour integral over the closed curve $C$ with the same value. That contour is deformed into the lower half of the $\omega$-plane encircling the eigenvalues $\omega = \omega_n$, so that the Cauchy residue theorem, or rather the Mittag-Leffler theorem, yields the value of the contour integral and, hence, the solution of the initial value problem for systems with discrete eigenvalues:
\BEQAR
E(x;t) &=& \sum_{\omega_n} {\rm e}^{-\hs{\rm i}\hs{\omega_n}t} \!\int_0^\infty {\rm Res} \big\{G(x,x';\omega)\big\}_{\omega_n} X(x';\omega_n) {\hs}d{\hs}x' \nonumber\\
&=& \sum_{\omega_n}  \hs{E(x;\omega_n)}\hs{\rm e}^{-\hs{\rm i}\hs{\omega_n}t} 
 \!\int_0^\infty E(x';\omega_n) {\hs}X(x';\omega_n) {\hs\hs}d{\hs}x' \,.\En{4.15}\EEQAR
 In the last equality, the expressions~\E{2.11} and \E{2.12} for $\Gamma$ and $\Delta$ have been inserted. This yields the standard solution of the conservative spectral problem when the continuum is approximated by a collection of closely packed discrete eigenvalues. The exact expression for the continuum contribution by an integral over the spectral cuts $\sigma \ge  k$ and $\sigma \le -k$ is omitted since the standard conservative picture is not our main concern here. Rather, we will demonstrate how the usual orthonormality conditions entering in the CNM eigenfunction expansions are modified for the QNMs.

\subsection{Orthogonality and normalisation}{\Sn{4.2}}

The final distinction between conservative and leaky modes comes from the different orthogonality relations of the modes and the different normalisations connected with them. In particular, the solution~\E{4.15} of the initial value problem does not make sense at all for the leaky modes because  the exponential explosion at infinity prevents expanding the constituent functions $E(x;\omega_n)$ of the Green's function in regular square integrable functions. In order to produce meaningful results, these functions need to be \IT{regularised}. This is the subject of sub-section~\S{4.2.2}.

General orthogonality and normalisation relations, both for conservative and for leaky modes, are obtained from Eqs.~\E{2.3} by considering two modes satisfying the respective equations for the frequencies $\omega_n$ and $\omega_m$:
\BEQAR
- {\rm i}\hs(\omega_n/v^2) \hs\bfE_n  &=& \curl{\bfH}_n \;\;(a)\,, \qquad - {\rm i}\hs\omega_n \hs\bfH_n  \hor{2}= - \curl{\bfE}_n \hor{2.5}(b)\,,\En{4.16}\\[2mm]
- {\rm i}\hs(\omega_m/v^2) \hs\bfE_m  &=& \curl{\bfH}_m \;\;(a)\,, \qquad - {\rm i}\hs\omega_m \hs\bfH _m = - \curl{\bfE}_m \;\;(b)\,.\En{4.17}\EEQAR
Quadratic forms are constructed from these equations by means of inner products, integrated over all space and reduced with the divergence theorem, schematically:
\BEQ \int \Big\{\hs\big[\hbox{\E{4.16}(a)} \cdot \,\bfE_m \,- \hbox{\E{4.17}(a)} \cdot \,\bfE_n\big] \,\pm \big[\hbox{\E{4.16}(b)} \cdot \,\bfH_m - \hbox{\E{4.17}(b)} \cdot \,\bfH_n\big]\hs\Big\} {\hs}dV\,, \non\EEQ
or explicitly:
\BEQ - {\rm i} \hs(\omega_n - \omega_m) \int_{\Omega^\pm} \hor{-1.5}\big(\bfE_n\cdot\bfE_m/v^2 \pm \bfH_n\cdot\bfH_m\big) \,dV = \pm \int_{\Sigma^\pm} \hor{-1.5}\big(\bfE_n \times \bfH_m - \bfE_m \times \bfH_n \big) \cdot \bfn \,dS \,. \En{4.18}\EEQ
Note that the arbitrariness in the choice of how the two magnetic evolution equations are folded in is exploited to produce a crucial difference between  two normalisations. In the two sub-sections below, we will show how the different signs correspond to different function spaces. For conservative modes (upper sign), the complete volume $\Omega^+$ is enclosed by the surface $\Sigma^+$, with functions evaluated far away from the dielectric (for our case, at $x = L_x \rightarrow \infty$). For leaky modes (lower sign), the restricted volume $\Omega^-$ is enclosed by the surface $\Sigma^-$, with functions evaluated just outside the region of inhomogeneity (for our case, at $x = 1$).

\subsubsection{Regular normalisation of the conservative modes}{\Sn{4.2.1}}

The regular normalisation of the conservative modes, corresponding to the choice of the $+$ sign in Eq.~\E{4.18}, consists of the \IT{orthogonality relation} for normal modes with different eigenvalues ($\omega_n \ne \omega_m$),
\BEQ \int_{\Omega^+} \hor{-.5}\big(\bfE_n\cdot\bfE_m/v^2 + \bfH_n\cdot\bfH_m\big) \,dV + \frac{\rm i}{\omega_m - \omega_n} \int_{\Sigma^+} \hor{-.5}\big(\bfE_n \times \bfH_m - \bfE_m \times \bfH_n \big) \cdot \bfn \,dS = 0 \,,\En{4.19}\EEQ
and the \IT{normalisation} of the eigenvalues (obtained from the limit $\omega = \omega_m \rightarrow \omega_n$),
\BEQ \int_{\Omega^+} \hor{-.5}\big(\bfE_n\cdot\bfE_n/v^2 + \bfH_n\cdot\bfH_n\big) \,dV + {\rm i} \int_{\Sigma^+} \hor{-.5}\Big(\bfE_n \times \frac{\partial\bfH_n}{\partial\omega} - \frac{\partial\bfE_n}{\partial\omega} \times \bfH_n \Big) \cdot \bfn \,dS = C_n^+ \En{4.20}\EEQ
with arbitrary normalisation constants $C_n^+$.
Here, as required by analytic continuation, similar to the expansion~\E{4.12} of the dispersion equation, the limit $\omega \rightarrow \omega_n$ is obtained by replacing the eigenvalue $\omega_m$ by an arbitrary frequency $\omega$. This is permitted since the equations~\E{4.16} and \E{4.17} are also valid for arbitrary frequencies.

We will complete the expression for the normalisation of the TE modes in the special configuration investigated in this paper. Restricting the homogeneous region to a finite range $1 \le |x| \le L_x$, and exploiting the expressions for {\em standing CNMs,} i.e.\ $|\sigma| \ge k$, $\hat{\alpha} = \hat{\beta} \exp(-2\hs{\rm i}\hs\varphi)$, and $\hat{q}_2 = 0$,  the  external solutions~\E{2.13} satisfying $\hat{E}_n(L_x) = 0$ yield the dispersion equation, $\hat{q} L_x + \varphi = (n + \half) \pi$, which depends on $\omega$ through the wave vector $\hat{q} = \pm \sqrt{\omega^2 - k^2}$ and through the angle $\varphi$ determined by $E(1)$ and $H(1)$. This gives
\BEQAR
&&\hat{E}_n(x) = 2\hs(-1)^n \hat{\beta}{\hs}e^{-{\rm i}\varphi} \sin\big(\hat{q}\hs(L_x - x)\big) \,,\non\\[2mm]
&&\hat{H}_n(x) = 2\hs{\rm i}\hs(-1)^n (\hat{q}/\omega)\hs\hat{\beta}{\hs}e^{-{\rm i}\varphi} \cos\big(\hat{q}\hs(L_x - x)\big)\,.\En{4.21}\EEQAR
Exploiting these equations, the external part $\hat{I}_{\Omega^+{\rm ext}}$ of the volume integral and the surface integral $\hat{I}_{\Sigma^+}$ of the normalisation~\E{4.20}  reduce to
\BEQ \hat{I}_{\Omega^+{\rm ext}} =  4 \hat{\alpha} \hs\hat{\beta} \hs(L_x - 1)\,,\qquad \hat{I}_{\Sigma^+} = 0 \,.\En{4.22}\EEQ
Consequently, the explicit normalisation condition for standing CNMs becomes
\BEQ \int_0^{1} \hor{-.5}\big[(1/v^2 + k^2/\sigma^2) E_n^2 + H_n^2\big] \,dx + 4 \hat{\alpha} \hat{\beta}\hs(L_x - 1) = C_n^+ \,,\En{4.23}\EEQ
where the size $L_x$ of the external region is finite for the discrete modes, and infinite for the dense continuum modes.

For the solution~\E{4.15} of the initial value problem, the eigenfunctions $E_n(x)$ are considered to be orthonormal, satisfying Eqs.~\E{4.19} and \E{4.20}. However, they have been normalised already above by prescribing the value $1$ to the quantity $A_n$ defined in Eq.~\E{4.12}. There is no conflict here, since the functions thus normalised will have a well-defined value of $C^+_n$ which, in our case, is given by computing the left hand side of Eq.~\E{4.23}. If one wishes to normalise the functions such that $C^+_n$ becomes unity, the quantity $A_n$ as defined in Eq.~\E{4.12} should be computed instead and inserted in the Green's function.

\subsubsection{Regularisation of the leaky modes}{\Sn{4.2.2}}

\IT{Regularisation} of the leaky modes is obtained by the choice of the $-$ sign in Eq.~\E{4.18}. This produces the \IT{orthogonality relation} (for $\omega_n \ne \omega_m$):
\BEQ \int_{\Omega^-} \hor{-.5}\big(\bfE_n\cdot\bfE_m/v^2 - \bfH_n\cdot\bfH_m\big) \,dV  - \frac{\rm i}{\omega_m - \omega_n} \int_{\Sigma^-} \hor{-.5}\big(\bfE_n \times \bfH_m - \bfE_m \times \bfH_n \big) \cdot \bfn \,dS = 0 \,,\En{4.24}\EEQ
and \IT{normalisation} of the eigenvalues (for $\omega = \omega_m \rightarrow \omega_n$):
\BEQ \int_{\Omega^-} \hor{-.5}\big(\bfE_n\cdot\bfE_n/v^2 - \bfH_n\cdot\bfH_n\big) \,dV - {\rm i} \int_{\Sigma^-} \hor{-.5}\Big(\bfE_n \times \frac{\partial\bfH_n}{\partial\omega} - \frac{\partial\bfE_n}{\partial\omega} \times \bfH_n \Big) \cdot \bfn \,dS = C_n^- \,.\En{4.25}\EEQ
The latter agrees with Eq.~(31) of \Ron{Sauvan2022}, where a more general expression is presented including effects that are of no interest for the present analysis. 

For the present case of leaky TE modes in the plane slab configuration, the electromagnetic fields are represented by 
\BEQ \hat{E}_n(x) = \hat{\alpha} e^{{\rm i} \hat{q} x}\,, \qquad \hat{H}_n(x) = (- {\rm i}/\omega) E' = (\hat{q}/\omega) \hs\hat{\alpha} e^{{\rm i} \hat{q} x} \,.\En{4.26}\EEQ
Hence, in the external region, the two fields have become proportional. This has  the far reaching consequence that  the volume integral over that region does not contribute to the normalisation, whereas the surface integral reduces to
\BEQ \hat{I}_{\Sigma^-} = - {\rm i}\hs({\partial\hat{q}}/{\partial\omega}) \big[\hat{E}_n\hs({\partial\hat{H}_n}/{\partial\hat{q}}) - ({\partial\hat{E}_n}/{\partial\hat{q}})\hs\hat{H}_n\big]_{x = 1} = - {\rm i}\hs[k^2/(\hat{q}\hs\omega^2)] \hs\hat{E}_n^2(1) \,. \En{4.27}\EEQ
Consequently, the normalisation for leaky QNMs only involves the contributions of the internal region and of its bounding surface (i.e.\ at $x = 1$!):
\BEQ \int_0^1 \hor{-.5}\big[(1/v^2 - k^2/\omega^2) E_n^2 - H_n^2\big] \,dx - {\rm i}\hs[k^2/(\hat{q}\hs\omega^2)] \hs\hat{E}_n^2(1) = C_n^- \,.\En{4.28}\EEQ
Note that the integrand of the volume integral~\E{4.23} for the CNMs consists of the sum of the electromagnetic energy densities of the two fields, whereas in the corresponding integral~\E{4.28}  the two energy densities are exactly cancelled in the external region for the leaky QNMs. Also note that the two operations of addition or subtraction of the two magnetic field equations~\E{4.16}(b) and \E{4.17}(b) and the two electric field equations~\E{4.16}(a) and \E{4.17}(a) are perfectly legal operations: both are valid, there are no approximations involved. Approximations do enter though when the expressions are applied to the initial value problem to determine the evolution of the system for times that are actually not long enough to apply the CNM formulation or not short enough to apply the QNM formulation. Numerical integration of the time-dependent equations for intermediate times should then give the connection between the two pictures.

\subsection{Summarising remarks on the leaky electromagnetic waves}{\Sn{4.3}}

The main point of the present section is the demonstration, taken from the extensive literature on electromagnetic waves in dielectric media (see e.g. \Ron{Sauvan2022} and references cited there), that the leaky modes can be put on a firm mathematical basis by a regularisation involving a norm alternative to the standard conservative norm. This justifies the terminology of distinguishing quasi-normal modes (QNMs) from conservative normal modes (CNMs). Of course, this extensive theoretical activity is dictated by the even more extensive literature on the experimental observation, and use, of leaky modes in dielectric media (see e.g. \Ron{Huang2023} and references cited there). The present section should serve as a benchmark for comparison with leaky magnetohydrodynamic waves in coronal magnetic flux tubes (our next topic) which frequently are also called quasi-normal modes and justified by assuming analogy with the electromagnetic wave problem. Given the uncertainty of what is an analogous configuration in the coronal plasma, of whether leaky modes are observed at all, and of the conflicting theoretical literature on this topic, it might serve to summarise the main assumptions underlying the theory of the quasi-normal modes in dielectric media.

A paper by \Ron{Ching1998} is quite helpful in this respect. It points out the wide range of applications of QNM resonances with complex frequencies in non-conservative systems described by non-Hermitian operators, from optics in microspheres, to gravitational waves from black holes, and even second-quantised versions of these theories. Based on earlier work of the same authors, completeness of the QNM expansion is proved provided two conditions are met: (1) {\em discontinuity} of the phase velocity at the boundary of the dielectric, (2) {\em no-tail} of the phase velocity beyond that boundary. The first condition is interpreted there as a jump of  a constant phase velocity to the velocity of light in vacuum. Presumably, the restriction to constant phase velocity in the inner region may be relaxed to admit arbitrary functions, as in the present paper, but the second condition is crucial. It ensures a clear distinction between an inner region with multiple scattering and an outer region with no back-scatter at all. This is nicely illustrated in Fig. 3 of their paper which shows a schematic space-time diagram of the Green's function in the different regions at different times. It demonstrates the distinct 
time scales of the conservative and  the leaky modes. In the next sections, we will investigate whether this may be translated into a similar analysis of MHD waves in coronal magnetic flux tubes.

\vfill\eject

\section{Magnetohydrodynamic waves in a magnetised plasma slab}{\Sn{5}} 

\subsection{Basic MHD model}{\Sn{5.1}}

In order to describe the analogy between electrodynamic waves in a dielectric medium and 
MHD waves in a magnetised plasma slab, it is imperative to first present the 
different assumptions and approximations underlying the spectral MHD equation that is to be 
compared with the `analogous' electrodynamic spectral problem~\E{2.3}. First of all, it needs 
to be stressed that the MHD waves considered here are linear perturbations of \IT{a 
magnetically confined plasma} in an equilibrium that is assumed to be static (no background 
flow). It would be described by the balance of the macroscopic Lorentz force and the pressure 
gradient,
\BEQ \bfj \times \bfB = \nabla p \,, \qquad \bfj = \curl{\bfB} \,,\En{5.1}\EEQ
where the factor $\mu_0$ is suppressed in anticipation of the natural units 
introduced below. For simplicity, we neglect all pressure and current effects so that 
Eq.~\E{5.1} is trivially satisfied (there is no magnetic confinement anymore, but there still is a 
magnetic `backbone' guiding the waves!). We keep a gradient of the density, $\rho = \rho(x)$, 
as the sole representative of a cold  flux tube in the solar corona. Since the magnetic 
field is assumed constant everywhere ($\bfj = 0$), the only background variable is the Alfv\'en velocity,
\BEQ b(x) \equiv B/\sqrt{\rho(x)} \,.\En{5.2}\EEQ
It is assumed constant outside (given by $\hat{b}$ for $|x| > 1$)
 but varying inside the plasma slab ($|x| \le 1$), very much 
like the phase velocity $v(x)$ of the electromagnetic waves in Eq.~\E{2.2}. Corresponding to the scaling~\E{2.1}, the natural units for the MHD problem are obtained from the following transformation:
\BEQ x/a \rightarrow x \,,\qquad t\hs(\hat{b}/a) \rightarrow t \,, \qquad b/\hat{b} \rightarrow b 
\,,\En{5.3}\EEQ
so that the scaled Alfv\'en speed $\hat{b}$ in the outer plasma has the value $1$ (like the scaled velocity of light in vacuum, $c = 1$, in the electromagnetic problem).

The MHD spectral problem, `Newton's law' for a plasma element, described in generality by Eqs.~(7.66)--(7.68) of~\Ron{GKP2019},  now reduces to
\BEQ \rho \deldeltt{\bfxi} = \bfF(\xi) \equiv \bfB \times [\curl{\curl{(\bfB \times \bfxi)}}]  
\quad\Rightarrow\quad (\oms/b^2){\bfxi} + \bfe_z \times [\curl{\curl{(\bfe_z \times \bfxi)}}]  
= 0\,.\En{5.4} \EEQ
Here, $\bfxi$ is the plasma displacement away from the equilibrium position and $\bfe_z$ is 
the unit vector in the direction of the magnetic field (chosen along the $z$-axis depicted in 
Fig.~\F{1}). Due to the scaling in natural units, the background equilibrium magnetic field $\bfB$ is now represented by the unit vector $\bfe_z$.
Although similar to the spectral problem~\E{2.3}, the MHD spectral problem at 
once reveals the crucial difference of the MHD waves: they are guided by the magnetic 
`carrier' whereas electromagnetic waves do not have a carrier. Nevertheless, 
electromagnetic waves are also guided, viz.\ \IT{engineered} to do so, e.g.\ along the length of an 
optical fiber. Apparently, the two spectral problems lead to very similar differential equations, 
suggesting the existence of leaky modes in both systems. We will have to investigate whether 
the mentioned physical differences between the two systems has any influence on this.

Note that Eq.~\E{5.4} implies
\BEQ \oms \bfB \cdot \bfxi = 0 \,,\En{5.5}\EEQ
so that, in general (for $\omega \ne 0$), there is no plasma displacement (no flow) along 
the magnetic field or, at marginal stability ($\omega = 0$), this flow has an arbitrary 
distribution not interacting with the other components. Hence, the sub-spectrum of slow magneto-acoustic waves has been eliminated (or rather,
 collapsed into the origin of the $\omega$-plane) and only the sub-spectra of Alfv\'en and fast 
magneto-acoustic waves remain. Similar to the electromagnetic wave spectrum, the MHD 
spectrum is then described by two transverse variables.

The central property of ideal (= dissipationless) magnetohydrodynamics is the conservation of magnetic flux, expressed by the relation between the displacement vector $\bfxi$ and the  magnetic field perturbation
\BEQ \bfQ = \curl{(\bfxi \times \bfB)} \,.\En{5.6}\EEQ
Associated with this variable is the perturbation of the total pressure,
\BEQ \Pi \equiv \bfB \cdot \bfQ = B Q_z \,,\En{5.7}\EEQ
which, in this case, only consists of the perturbation of the magnetic pressure. The MHD spectral problem turns out to be described completely by the displacement $\bfxi \cdot \bfn$ normal to the magnetic surfaces and the magnetic pressure $\Pi$, both considered as functions of the coordinate labelling the magnetic surfaces. Here, since there is no equilibrium current, the magnetic surfaces will be the surfaces of constant $x$.

\subsection{Ordinary differential equations for Alfv\'en and fast magnetosonic waves}{\Sn{5.2}}

We again exploit the simple configuration of Fig.~\F{1} since it is also exemplary for   
MHD waves in magnetised plasmas. For the same reason as in the electromagnetic case, the one-dimensional MHD spectral
 problem may be described by normal modes of the form~\E{2.6}. In contrast to the sequel in Sec.~\S{2.2}, the
 mode number $\ell$ in the $y$-direction is now kept producing the following representation of the spectral equation~\E{5.4}:
\BEQ \left(\begin{array}{@{}c@{\quad}c@{}}
\partial_x^2 + \omega^2/b^2 - k^2 & {\rm i} \hs\ell \hs\partial_x\\[2mm]
 {\rm i} \hs\ell \hs\partial_x & \omega^2/b^2 - (k^2 + \ell^2) 
\end{array} \right) 
\left(\begin{array}{@{}c}
\xi_x \\[2mm]
\xi_y 
\end{array} \!\!\!\right) = 0 \,. \En{5.8}\EEQ
According to Eq.~\E{5.5},
\BEQ \oms \xi_z = 0 \,,\En{5.9}\EEQ
so that $\xi_z \equiv 0$ for $\omega \ne 0$, whereas $\xi_z(x)$ is arbitrary with $\xi_x = \xi_y = 0$ for $\omega = 0\,$. 

Eliminating the component $\xi_y$, the system \E{5.8} is reduced to one second order ODE:
\BEQ \ddx{}\hs\bigg(\frac{\omega^2/b^2 - k^2}{\omega^2/b^2 - k^2 - \ell^2} \,\ddx{\xi}\bigg) + (\omega^2/b^2 - k^2)\hs \xi = 0 \,,\quad\hbox{where}\;\; \xi \equiv \xi_x \,.\En{5.10}\EEQ
Equivalently, by exploiting the magnetic pressure perturbation $\Pi = Q_z = -\xi'_x - {\rm i}\hs\ell \hs\xi_y$, and again eliminating $\xi_y$, the equivalent system of two first order ODEs becomes:
\BEQ
\left\{\;\begin{array}{@{}l@{}l}
\frac{\ds d \hs\hs\xi}{\ds {\hs}dx} + \frac{\ds\omega^2/b^2 - k^2 - \ell^2}{\ds\omega^2/b^2 - k^2} \Pi = 0 \,, \\[3mm]
\frac{\ds d \hs\Pi}{\ds {\hs}dx} - (\omega^2/b^2 - k^2) \xi = 0 \,.
\end{array} \right. \En{5.11}\EEQ
The second order ODE is most convenient for explicit analysis, e.g.\ of the solutions in the external plasma (where $b = \hat{b} \equiv 1$), whereas the system of first order ODE is most adequate for numerical solution in the internal plasma (where $b = b(x)$). Evidently, solution of the BVP by matching the internal to the external solution at $|x| = 1$ will involve both. It will be noticed that the first order system~\E{5.11} for the $\ell = 0$ fast modes becomes identical to the first order system~\E{2.11} for the TE modes when the following identifications are made:
\BEQ E \;\rightarrow\; v \equiv \partial_t \xi = - {\rm i} \hs\omega \hs\xi \,,\qquad H \;\rightarrow\; \Pi \,,\En{5.12}\EEQ
where $v$ is the velocity perturbation in the $x$-direction. This is convenient for checking the numerical procedures. Clearly, these identifications no longer hold for $\ell \ne 0$, when the Alfv\'en waves enter the picture. 

At this point, it should be stressed that the second order ODE~\E{5.10} for MHD waves in plane geometry is, in fact, a gross simplification of the dynamics of these waves in actual magnetic flux tubes observed in the solar corona. However, for the more realistic cylindrical geometry and arbitrary equilibrium distributions, the analogous equation derived long ago by~\Ron{HL58}, exhibits the same structure with a singular factor in front of the highest derivative (albeit slightly more general, including the singular contributions of the slow magneto-sonic modes). Possible instabilities due to pressure and current contributions are only represented by additional terms in front of the last term, the one multiplying $\xi$ in Eq.~\E{5.10}. In particular, the fundamental paper by~\Ron{Newcomb1960} on the stability of all diffuse linear pinches is obtained in the limit $\omega \rightarrow 0$ of this equation. Hence, as stated above, the present plane geometry may be considered as sufficiently exemplary to investigate the physical reality of leaky fast magneto-sonic modes in coronal flux tubes.

There is yet another matter that should be stressed: the innocuously looking elimination of $\xi_y$ from the spectral problem does not have a counterpart in the dielectric problem. In MHD, that variable describes the plasma displacement tangential to the magnetic surfaces, which is the main component of the Alfv\'en waves, i.e.\ of the central constituents of MHD spectra! It becomes singular for frequencies $\omega \rightarrow \omega_{\rm A} = k b(x_{\rm s})$ at positions $x = x_{\rm s}$, as is evident from the second relation~\E{5.8} for $\ell = 0$:
\BEQ (\omega^2/b^2 - k^2) \hs\xi_y = 0 \quad\Rightarrow \quad \xi_y = C \hs\delta(x - x_{\rm s}) \,,\En{5.13}\EEQ
whereas the function $\xi_x(x)$ has no influence on this. For $\ell \ne 0$, the algebra becomes a little more involved with Heaviside functions entering the solutions $\xi_x$ of the ODE~\E{5.10}, but the resulting component $\xi_y$ remains the dominant $\delta$-function contributor.

The merit of keeping the mode number $\ell$ in the ODE~\E{5.10} is that the genuine Alfv\'en singularities in the numerator of the term multiplying the second derivative, at $\omega = \pm \omega_{\rm A}(x) \equiv \pm k b(x)$, are not cancelled by the `fake' fast singularities in the denominator, at $\omega = \pm \omega_{\rm f0}(x) \equiv \pm k_{\rm tot}{\hs}b(x)$ with $k_{\rm tot} \equiv \sqrt{k^2 + \ell^2}$. The Alfv\'en singularities produce the \IT{continuous Alfv\'en spectra,}
\BEQ \omega_{\rm A,\,min} \;\big\{\!= k b(1 - \epsilon) \equiv k \delta \hs\big\}\; \le \;|\sigma| \le \omega_{\rm A,\,max} \;\big\{\!= k b(1) \equiv k\hs\big\} \,,\En{5.14}\EEQ
corresponding to `improper' eigenfunctions which are singular for the real frequencies $|\sigma| = \omega_{\rm A}(x_{\rm s})$ in the range $1 - \epsilon \le x_{\rm s} \le 1$. The explicit expressions in curly brackets for the boundaries of this continuum are obtained from the choice~\E{3.1} for the phase velocity replacing $v(x)$ by $b(x)$. The \IT{fast apparent singularities} $\omega = \pm \omega_{\rm f0}(x)$ just indicate where the monotonic relation between frequency and oscillations of the eigenfunctions no longer holds~\R{GS1974}, although the eigenfunctions themselves remain quadratically integrable. In general, the oscillatory properties of the sub-spectra change at the boundaries of both the singular and  the apparently singular frequency regions. Hence, a Sturmian sequence of discrete Alfv\'en waves could occur for frequencies below the lower edge of the Alfv\'en continuum, $0 < |\sigma| < \omega_{\rm A,\,min}$, and an anti-Sturmian sequence of discrete Alfv\'en waves above the upper edge, $\omega_{\rm A,\,max} < |\sigma|  < k_{\rm tot}$. Because of the extremely simplified present configuration, there are no discrete Alfv\'en waves in this problem but only Alf\'en continuum modes. However, there is a sequence of discrete fast waves which is Sturmian for $k_{\rm tot} < |\sigma| < \infty$, that would cluster for $|\sigma| \rightarrow \infty$ but only when the external plasma is finite. In the present case, it is replaced by a continuum since the external plasma is assumed to be infinite. In addition, fast discrete modes may occur, and here actually do occur, for $\omega_{\rm A,\,max} <  |\sigma| \le k_{\rm tot}$. That sequence could connect to an anti-Sturmian sequence of discrete Alfv\'en waves if those would be permitted (hence, the proof of monotonicIty fails in that frequency range). All this only refers to conservative modes, the leaky ones will be studied in Sec.~\S{6}.

Since the Alfv\'en speed is constant in the external plasma, the two factors in the ODE~\E{5.10} can be extracted from the derivative for $|x| > 1$ to yield the external dispersion equation of the Alfv\'en and fast magneto-sonic modes,
\BEQ (\omega^2 - k^2)(\omega^2 - \hat{q}^2 - k^2 - \ell^2) = 0 \,,\En{5.15}\EEQ
where $\hat{q}$ is the mode number in the $x$-direction appearing in the expression for the external variables $\hat{\xi}(x)$ and $\hat{\Pi}(x)$, analogous to Eq.~\E{2.13}. Since the first factor exclusively vanishes for real values of $\omega$ in the ranges of the Alfv\'en continua, inversion of the external dispersion equation to produce the relation $\hat{q} = \hat{q}(\omega)$ needed for the construction of the Spectral Web is nearly the same as in Sec.~\S{2.3} for the TE modes. In particular, the explicit solution~\E{2.14} and Fig.~\F{2} delineating the different cutoff and propagation regions remain valid, provided that, on the real axis, the crossings at $|\sigma| = k$ are replaced by $|\sigma| = k_{\rm tot}$ and the isolated strips $k \delta \le |\sigma| \le k$ are excluded.

As an aside, the spectral equation~\E{5.10} was studied before in one of the first papers on Alfv\'en wave heating in tokamaks by~\Ron{ChenHasegawa1974}, extensively described in Section (11.1) of~\Ron{GKP2019}. As customary for laboratory settings, instead of external fast modes, an external vacuum with exciting  electromagnetic waves was assumed. Corresponding to the non-relativistic MHD model, their frequency could be neglected, but they effectively couple to the internal fast waves and then, through wave transformation, to the singular Alfv\'en waves which would be `dissipated' by continuum damping. Clearly, this mechanism should also work in solar magnetic flux tubes with the present fast external waves as an exciting mechanism. It is also clear now that this requires fast wave frequencies in the cutoff region, where the Alfv\'en waves are located. The present paper is not concerned with Alfv\'en wave heating but with the physics of leaky modes, which occur for fast wave frequencies in the propagating region. As a further aside, it should be mentioned that the replacement of a vacuum magnetic perturbation (customary for fusion experiments) by a plasma with fast magnetosonic waves in the more realistic cylindrical geometry for flux tubes requires the analysis of Bessel or Hankel functions with complex argument, like ${\rm J}_m(\sqrt{\omega^2 - k^2}{\hs}r)$. This complication is avoided here by exploiting a plane slab geometry permitting straightforward inversion of the external dispersion equation.

\subsection{Boundary value problem for Alfv\'en--fast magnetosonic waves}{\Sn{5.3}} 

Analogous to Sec.~\S{2.3} on the TE modes, the internal, numerical, solutions of the ODEs~\E{5.11} for the Alfv\'en--fast magnetosonic spectral problem are written as
\BEQ \xi = \xi(x) \,,\quad \Pi = \Pi(x) \qquad\hbox{($|x| \le 
1$)} \,,\En{5.16}\EEQ
whereas the same ODEs, for $b(x) = \hat{b} \equiv 1$ and eliminating $\omega$ with the external dispersion equation~\E{5.15}, provide the explicit external solutions:
\BEQ \hat{\xi}(x) = \hat{\alpha}\hs{\rm e}^{{\rm i}\hat{q}x} + \hat{\beta}\hs{\rm e}^{-{\rm i} 
\hat{q}x} \,,\quad \hat{\Pi}(x) = - {\rm i}\hs\hat{q} \hs(1 + \ell^2/\hat{q}^2)(\hat{\alpha}\hs{\rm e}^{{\rm i}\hat{q}x} - \hat{\beta}\hs{\rm e}^{-{\rm i} 
\hat{q}x}) \qquad\hbox{($|x| > 1$)} \,. \En{5.17}\EEQ
The  boundary conditions to match these solutions are the same as Eqs.~\E{2.23}--\E{2.25}, of course with proper changes of the variables: 
\BEQAR
\hor{-10}&&\hbox{(a)~at $|x| = 0$:} \quad \xi(0) = 0 \quad\hbox{\IT{(odd modes)}}\,, \quad 
\hbox{or}\quad \Pi(0) = 0 \quad\hbox{\IT{(even modes)}}  \,,\En{5.18}\\[2mm]
\hor{-10}&&\hbox{(b)~at $|x| = 1$:} \quad \xi(1) = \hat{\xi}(1) \;\; \hbox{and}\;\; \Pi(1) = 
\hat{\Pi}(1) \,,\En{5.19}\\[2mm]
\hor{-10}&&\hbox{(c)~at $|x| \rightarrow \infty$\,,}\; 
\left\{\begin{array}{@{}l@{}l}
\,(c1)\; \hbox{\IT{CNMs}:}  \;\;\hat{\alpha} = 0 \;\;\hbox{(for $|\sigma| < k_{\rm tot}$)}\,,\quad |\hat{\alpha}| = |\hat{\beta}| \;\;\hbox{(for $|\sigma| \ge k_{\rm tot}$)}\,,\\[4mm]
\,(c2)\; \hbox{\IT{Leaky modes}:} \;\;\hat{\beta} = 0\,.
\end{array} \right. \En{5.20}\EEQAR
Here, the auxiliary parameter $k_{\rm tot} \equiv \sqrt{k^2 + \ell^2}$ marks the cutoff boundary, and the leaky fast magnetosonic modes are not indicated as QNMs since it is to be proved yet that this terminology would be appropriate for these modes.

\subsection{Spectral Web for magnetohydrodynamic waves  in magnetised plasmas}{\Sn{5.4}}

Justification of the above boundary conditions and construction of the Spectral Web for this case requires a quantity corresponding to the Poynting flux of Sec.~\S{2.4}. For this purpose, the potential energy $W$ of the perturbations will be chosen. It is a quadratic function of the displacement vector $\bfxi$ appearing in the force operator $\bfF$, which has been shown to be self-adjoint by rather cumbersome vector operations converting the original volume integral into a symmetric volume integral over the plasma complemented with a surface integral over the outer boundary. Since all this is standard theory, we just adapt the well-known expression to our system of the two plasmas by splitting the complete volume integral into two parts, where the crucial new construct is an additional surface integral over the separating boundary $S_1$ (at $|x| = 1$ for the present case):
\BEQAR W &\equiv& - \half \int \bfxi^* \cdot \bfF(\bfxi) {\hs}dV \non\\
&=& \half \int \bfQ\cdot\bfQ^* dV - \half \int \bfxi^*\!\cdot\bfn \hs\hs\doublel\Pi\doubler {\hs\hs}dS_1 + \half \int \hat{\bfQ}\cdot\hat{\bfQ}^* d\hat{V} + \int \hat{\bfxi}^*\!\cdot\bfn \hs\hs\hat{\Pi} {\hs\hs}d\hat{S} \,.\En{5.21}\EEQAR
The volume integrals only contain the magnetic field perturbations $\bfQ$ and $\hat{\bfQ}$ since the equilibrium current and pressure are assumed to vanish. The surface integrals are completely general. They manifest the surprising property of ideal MHD that all perturbations can be reduced to the two variables $\xi \equiv \bfxi \cdot \bfn$ and $\Pi$ which are propagated from magnetic surface to magnetic surface by the pertinent ODEs or PDEs. The second surface integral imports the outer BC, which may be situated at infinity. 

Corresponding to the definition~\E{2.29} of the complementary Poynting flux, averaged over a unit surface area, the first surface integral is the \IT{complementary energy},
\BEQ W_{\rm com} \equiv - \frac{1}{2 L_y L_z} \int_{-L_y}^{L_y} \int_{-L_z}^{L_z} 
\doublel{\Pi}\doubler \hs\hs\xi^*{\hs}dy{\hs}dz = - \half \hs\xi^*(1) \hs\doublel{\Pi(1)}\doubler \,.\En{5.22}\EEQ
It will be used to construct the Spectral web for arbitrary complex values of $\omega$ by computing internal solutions satisfying the BCs~\E{5.18}, external solutions satisfying either one of the two BCs~\E{5.20}, and matching them by imposing continuity of $\xi$ at $x = 1$. The second BC~\E{5.19}, continuity of $\Pi$, will then not be satisfied in general, except for the wanted eigenvalues. The second surface integral The last term of Eq.~\E{5.21}, the surface integral over the boundary at infinity, represents the energy that is being lost there (for the leaky modes) or not (for the conservative modes). The Spectral Web method to construct these two types of modes is completely analogous to the one used in Section~\S{2.4} to construct the spectra for the conservative and leaky TE modes so that we can now proceed to the similar computation for the Alfv\'en and fast magneto-sonic modes.

\section{Spectra of conservative and leaky Alfv\'en--fast modes}{\Sn{6}} 

\subsection{Preliminaries}{\Sn{6.1}}

Analogous to Sec.~\S{3.1} for the electromagnetic waves, we exploit the distribution~\E{3.1}, with the same two parameters $\epsilon$ and $\delta$ determining  size and strength of the inhomogeneity, whereas the phase velocity $v(x)$ is replaced by the Alfv\'en velocity $b(x)$. Now, an additional wave number $\ell$ in the $y$-direction enters so that the total wavenumber, tangential to the magnetic surfaces, becomes $k_{\rm tot} \equiv \sqrt{k^2 + \ell^2}$. Hence, in contrast to the three-parameter numerical code DLEAK (employed in Sec.~\S{3}) for the EM waves in dielectric media, the numerical code CLEAK (employed in this section)  for MHD waves in coronal loops contains four free parameters, viz.\ $\epsilon$, $\delta$ , $k$ and $\ell$. The value of $\epsilon$ will now be chosen larger to make the configuration for coronal loops one step more realistic. The values of the other parameters will be chosen largely for the purpose of illustration of the spectral characteristics since it is obvious that the configuration is a `caricature' anyway of the actual magnetic loops that are observed in the solar corona. Nevertheless, the calculation is valuable because it exhibits the main features of the intricate interplay of fast and Alfv\'en waves that recur in more realistic configurations.

Recall that the `wave number' $q$ in the direction of inhomogeneity is not a valid quantum number for the interior region $|x| \le 1$. Its local value is just a result of the numerical solution of the pertinent ODEs. The wave number $\hat{q}$ for the external region $|x| > 1$ is a valid quantum number though, determined by the inverse $\hat{q}(\omega)$ of the fast wave dispersion equation \E{5.15}. The Spectral Web method hinges on computing the solution paths and the conjugate paths for prescribed complex values of $\omega$. Hence, it is very fortunate that this inversion can be performed algebraically for this configuration, see Eq.~\E{2.14} and Fig.~\F{2} with $k$ replaced by $k_{\rm tot}$. Generalisation to cylindrical or more general geometries would be more complicated but, as mentioned at the end of Sec.~\S{3.4}, the external solution could then be obtained with the same numerical integration method as exploited for the interior region, i.e.\ for prescribed values of $\omega$. With the speed and accuracy of present numerical solvers, one needs not to bother anymore about computing zeros of Hankel functions,

\subsection{Conservative Alfv\'en--fast modes}{\Sn{6.2}} 

Representative spectra of the Conservative Normal Modes, i.e.\ the regular MHD waves in a coronal magnetic flux tube are depicted in Fig.~\F{7}. Compared to the EM spectra depicted in Figs.~\F{3} and \F{4}, there are now the additional contributions of the Alfv\'en waves in the MHD spectra of Figs.~\F{7} and \F{8}. To exhibit those most clearly, only the positive frequency parts of the spectra are dsplayed.

To understand the fundamental difference between the EM waves of Fig.~\F{3} and the MHD waves of Fig.~\F{7}, it may be helpful to recall the elementary properties of Alfv\'en  and fast waves for homogeneous plasmas. In that case, since $\xi_x \sim \xi_y \sim \exp({\rm i} q x)$, the ODEs~\E{5.8} become an algebraic system having, not surprisingly, the discriminant~\E{5.15} with $\hat{q}$ replaced by $q$. The solutions are:
\BEQAR
&&\hbox{-- {\em degenerate Alfv\'en waves} at $\hs\omega^2/b^2 = k^2\hs$, with $\hs\bfxi_{\rm A} \perp \bfk_\perp \equiv (q, \ell, 0)$}\,,\\[2mm]\En{6.1}
&&\hbox{-- {\em fast magneto-sonic waves} at $\hs\omega^2/b^2 = k_{\rm tot}^2 + q^2\hs$, with $\hs\bfxi_{\rm F} \parallel \bfk_\perp$}\,.\En{6.2}\EEQAR
For a finite box, $q$ becomes quantised and the fast mode spectrum becomes a discrete set of eigenvalues starting at $\omega^2/b^2 = k_{\rm tot}^2$. If the box is infinite, the fast spectrum becomes a continuum. For the present inhomogeneous case (extended with an external region with wave number $\hat{q}$) the local internal wave number $q(x)$ may become imaginary, giving rise to an extension with bound fast waves in the cutoff region $0 < |\sigma|/b < k_{\rm tot}$. With an inhomogeneous internal region, the degenerate  Alfv\'en spectrum also becomes a continuum with frequencies given by Eq.~\E{5.14}. (Keep in mind though that this continuum is of a completely different nature than the fast continuum.) A good fraction of the bound fast waves are `swallowed' by the Alfv\'en continuum. Whereas for $\ell = 0$ the fast mode spectrum is governed by the same ODE~\E{5.10} for $\xi_x$ as the ODE~\E{2.10} for $E_y$ governing the TE mode spectrum, it is obvious though that this analogy completely breaks down for $\ell \ne 0$ since EM waves do not have the similar counterpart of Alfv\'en waves. All this is clearly demonstrated by Fig.~\F{7}.

\subsection{Leaky Alfv\'en--fast modes}{\Sn{6.3}} 

Representative spectra of the leaky MHD waves in a coronal magnetic flux tube are depicted in Fig.~\F{8}. The apparent similarity (in particular in the propagating region) between the conservative TE mode spectra and the conservative MHD mode spectra, depicted in Figs.~\F{3} and \F{7}, and between the leaky mode spectra depicted in Figs.~\F{4} and \F{8}, demands a careful analysis though. In Fig.~\F{8}. a good fraction of the leaky MHD modes in the cutoff region is again `swallowed' by the Alfv\'en continuum, but the two leaky modes close to $\omega = 0$, at $\nu = 4.7048\times10^{-3}$ (even modes) and at $\sigma = 4.7070\times10^{-3}$ (odd modes) are obviously not affected by the Alfv\'en continuum. From Eq.~\E{5.10} these `solutions' already grow explosively as $\xi \sim \exp{(k_{\rm tot} |x|)}$ close to $|x| = 0$ so that they formally belong to the leaky mode variety. If we label them by $n = 1$, it already becomes complicated to recognise the connection with the other `eigenvalues' of the leaky modes in the cutoff region. For the particular choice of the parameters $k$ and $\ell$ in Fig.~\F{8} only two have remained, viz.\ $n = 7, 8$ (even modes) and $n = 6, 7$ (odd modes). For MHD, not much remains of the alternating sequences of conservative and leaky modes illustrated in Table \T{1}(a), i.e.\ not much of a spectral structure if one wishes to develop `coronal seismology' based on leaky fast modes! Here, we have just skipped the smaller mode labels and continued counting beyond the Alfv\'en continuum by following the same procedure as for the leaky TE modes depicted in Fig.~\F{4}, viz.\ by {\em counting the zeros} of the real part of the `eigenfunctions' {\em on the internal part $|x| \le 1$ of the configuration.}

In the propagating region, this counting procedure becomes even more dubious. The zeros of the real parts of the `eigenfunctions' simply jump one digit, whereas the imaginary parts do not  exhibit this jump (in contrast to the leaky TE mode spectra depicted in Fig.~\F{4} where the number of zeros of both the real and the imaginary parts is just continuous). This counting anomaly appears to be connected with a more serious shortcoming of the leaky MHD spectrum in the propagating region depicted in Fig.~\F{8}. Whereas the separate QNM `eigenvalues' of the electromagnetic problem shown in Fig~\F{4} are connected by the solution path (the red curve), which justifies the counting procedure based on the monotonicity theorems along the Spectral Web paths proved earlier~\R{Goed2018}, no such connection exists in the MHD problem! On the contrary, a blow up of the spectrum shows that both paths of the Spectral Web are `blocked' at the cutoff frequency $\sigma = k_{\rm tot}$, where the solution path (red) starts to wind about the cutoff curve (green) and then connects onto a conjugate path (blue) which abruptly ends in the propagating region, without providing a connection between the `eigenvalues' of the leaky mode sequence. Each one of them is found at the intersection of a closed loop of the solution path and a similar closed loop of the conjugate path: a beautiful structure, mathematically correct but the connection to the quadratic forms of a physically meaningful spectral theory is questionable. This brings us to the subject of the next section.



\section{Attempt to regularise the leaky fast MHD modes}{\Sn{7}} 

Next, the excitation of the MHD spectrum of Alfv\'en and fast waves in the solution of the initial value problem needs to be discussed, in analogy with Sec. \S{4} for the electromagnetic problem. The algebra of Sec.~\S{4.1}  for the corresponding MHD problem can be skipped because it is identical. One only has to replace the variable $E(x)$, satisfying the ODE~\E{2.10} by the variable $\xi(x)$, satisfying the ODE~\E{5.10}, and the associated variables $H(x)$ and $\Pi(x)$ according to the scheme~\E{5.12}.

Hence, it appears to remain to derive orthogonality and normalisation relations justifying the use of leaky fast modes in the solution of an initial value problem for coronal magnetic flux tubes, in the same way as in Sec.~\S{4.2} for TE modes in dielectric media. To that end, we start from the MHD time evolution equations in the primitive first order variables $\bfv$ and~$\bfQ$, since the perturbations of $\rho$ and $p$ do not contribute in this problem:
\BEQ \rho\deldelt{\bfv}  = - \bfB \times (\curl{\bfQ}) \,,\qquad \deldelt{\bfQ}  = - \curl{(\bfB \times \bfv)} \,.\En{7.1}\EEQ
These equations are to be considered as the counterpart of the basic equations~\E{2.3} for the electrodynamical variables $\bfE$ and $\bfH$ resulting in Eqs.~\E{4.16} and \E{4.17} of Sec.~\S{4.2}. Similarly, we here consider two modes satisfying the equations~\E{7.1} for the frequencies $\omega_n$ and $\omega_m$:
\BEQAR
- {\rm i}\hs(\omega_n/b^2) \hs\bfv_n  &=& - \bfB \times (\curl{\bfQ_n}) \;\;(a)\,, \qquad\hs - {\rm i}\hs\omega_n \hs\bfQ_n  \hs\hs= - \curl{(\bfB \times \bfv_n)} \;\;(b)\,,\En{7.2}\\[2mm]
- {\rm i}\hs(\omega_m/b^2) \hs\bfv_m  &=& - \bfB \times (\curl{\bfQ_m})  \;\;(a)\,, \qquad - {\rm i}\hs\omega_m \hs\bfQ _m = - \curl{(\bfB \times \bfv_m)} \;\;(b)\,.\En{7.3}\EEQAR
Quadratic forms are constructed from these equations according to the same scheme as exploited to get Eq.~\E{4.18}, where now the Alfv\'en speed $b \equiv B/\sqrt{\rho}$ will enter by dividing $\bfB$ and $\bfQ$ by the constant value $B$ of the background magnetic field, 
\BEQ \int_\Omega \Big\{\hs\big[\hbox{\E{7.2}(a)} \cdot \,\bfv_m \,- \hbox{\E{7.3}(a)} \cdot \,\bfv_n\big] \,\pm \big[\hbox{\E{7.2}(b)} \cdot \,\bfQ_m - \hbox{\E{7.3}(b)} \cdot \,\bfQ_n\big]\hs\Big\} {\hs}dV \,. \non\EEQ
This yields two quadratic forms, analogous to the electromagnetic problem:
\BEQAR
&& - {\rm i} (\omega_n - \omega_m) \int_\Omega \!\big(\bfv_n\cdot\bfv_m/b^2 \pm \bfQ_n\cdot\bfQ_m\big) \,dV \non\\
&&\qquad= \int_\Omega \big[- \bfv_m \cdot \big(\bfB \times (\curl{\bfQ_n})\big) + \bfv_n \cdot \big(\bfB \times (\curl{\bfQ_m})\big)\big] \,dV
 \non\\[-2mm]
&&\qquad\qquad\qquad\pm \int_\Omega \big[- \bfQ_m \cdot \big(\curl{(\bfB \times \bfv_n)}\big) + \bfQ_n \cdot \big(\curl{(\bfB \times \bfv_m)}\big)\big] \,dV
 \non\\
 &&\qquad= - (1 \pm 1) \int_\Omega \big[\hs\bfQ_n\cdot\bfR_m - \bfQ_m\cdot\bfR_n\big] \,dV - \int_\Sigma \big[\hs\bfB\cdot\bfQ_n \bfv_m -  \bfB\cdot\bfQ_m \bfv_n\big] \cdot \bfn \,dS \non\\
 &&\qquad\equiv - (1 \pm 1)\hs\hs{\rm i}\hs\hs(\omega_n - \omega_m) \int_\Omega \bfQ_n\cdot\bfQ_m \,dV - \int_\Sigma \big[\hs\Pi_n \bfv_m -  \Pi_m \bfv_n\big] \cdot \bfn \,dS \,.\En{7.4}\EEQAR
 Here, the reduction of the RHS volume integrals to surface integrals, as in Eq.~\E{4.18} for the electromagnetic problem, involved the following transformations: The first RHS volume integral was converted into a volume integral plus a surface integral by introducing the auxiliary variable $\bfR \equiv - \curl{(\bfB \times \bfv)}$. This involved exploiting the rather complicated algebraic manipulations that are common in the proof of self-adjointness of the MHD force operator, see~\Ron{GKP2019}, Eq.~(6.52). (Notice: self-adjointness itself was {\em not} exploited,  just the occurring vector identities!) In the last step. it was observed that the variable $\bfR_{n,m}$ is directly related to $\bfQ_{n,m}$,
\BEQ \bfR_{n,m} \equiv - \curl{(\bfB \times \bfv_{n,m})} = - {\rm i}\hs\hs\omega_{n,m}\hs\bfQ_{n,m} \,,\En{7.5}\EEQ
which appears to complete the construction of two quadratic forms \E{7.4} for the MHD problem, analogous to the EM problem. 

The very last step is an anti-climax though. It consists of moving the final RHS volume integral to the LHS to discover that it nullifies the options suggested by the LHS volume integral. Consequently, the equality~\E{7.4} collapses into a single expression for the MHD \IT{orthogonality relation:}
\BEQ - {\rm i} (\omega_n - \omega_m) \int_\Omega \big(\bfv_n\cdot\bfv_m/b^2 - \bfQ_n\cdot\bfQ_m\big) \,dV 
+ \int_\Sigma  \big[\hs\Pi_n \bfv_m -  \Pi_m \bfv_n\big] \cdot \bfn \,dS = 0 \,.\En{7.6}\EEQ
Hence, the associated \IT{normalisation} also collapsed into a single expression:
\BEQ \int_\Omega \big(\bfv_n\cdot\bfv_n/b^2 - \bfQ_n\cdot\bfQ_n\big) \,dV 
- {\rm i} \int_\Sigma \big[\hs\Pi_n \frac{\partial\bfv_n}{\partial\omega} - \frac{\partial\Pi_n}{\partial\omega} \hs\bfv_n\big] \cdot \bfn \,dS = C_n \,.\En{7.7}\EEQ
The volume integrals are now taken over the full domain $\Omega$, enclosed by the surface $\Sigma$ far away from the loop, or at infinity, so that the surface integrals either vanish due to the BC $\bfv \cdot \bfn = 0$, or contribute to the integral over the infinite domain associated with the fast wave continuum. Needless to say, all this refers to the conservative fast modes since the leaky fast modes have now been disqualified.

Not accidentally, the relation \E{7.5} directly refers to the connection~\E{5.6} between the magnetic field perturbation $\bfQ$ and the displacement vector $\bfxi$: \IT{plasma and magnetic field move together,} both inside the magnetic loop and outside in the surrounding plasma. In contrast, for the electromagnetic waves in the compound dielectric-`vacuum' system, such a constraint between the fields $\bfE$ and $\bfH$ does not exist. This difference between the two systems is clearly manifested by the differential equations~\E{2.3} for the dielectric system and the differential equations~\E{7.1} for the MHD system. In particular, the dynamical relation between $\bfE$ and $\bfH$ for the dielectric system becomes completely symmetric in the external `vacuum' when $v/c \rightarrow 1$. Consequently, the two terms on the LHS of Eq.~\E{4.18} exactly cancel for running waves in the vacuum. This results in the amazing consequence that the leaky TE modes can be regularised by a normalisation where the external volume integral does not contribute at all, as expressed by Eq.~\E{4.28}. The dynamical relation between $\bfv$ and $\bfQ$ for the MHD system lacks such a symmetry through the conspicuous presence of an additional cross product involving the background equilibrium magnetic field $\bfB$ in the dynamical equations. Equally amazing, while the two actions of adding and subtracting the dynamical equation pairs are perfectly permitted for both systems, the consequence of magnetic flux conservation eliminates the freedom for the MHD system corresponding to that of the dielectric system: both signs in Eq.~\E{7.4} refer to the single possibility of only normalising the conservative MHD modes. \IT{The leaky fast magneto-sonic waves cannot be regularised!} 

\vfill\eject

\section{Conclusions}{\Sn{8}} 

We have compared the spectral theories of leaky waves in two representative settings, viz.\ electromagnetic (EM) waves in dielectric media and magnetohydrodynamic (MHD) waves in coronal magnetic flux tubes, both in the simplest configuration of the plane slab geometry that has been current in the two separate research communities. The aim was to find out whether the two theories are analogous. The answer is: they are not! Common to both is the propagation of waves in an infinite homogeneous medium, without any back reflection of either a wall or inhomogeneities. Quite different is that ``free'' EM waves can well be envisioned as propagating in empty space (even air is reasonably well approximated), whereas there is no such a thing as ``free'' MHD waves: they need a supporting magnetised plasma ensuring conservation of the magnetic flux through all perturbed structures. The consequence is that the spectral theory of EM waves permits the construction of {\em two} quadratic forms for the normalisation of the perturbations, viz. the regular conservative one and a quasi-normal one singling out the effects of the interior inhomogeneity of the dielectric material. Spectral theory of MHD waves, on the other hand, only permits {\em one} normalisation, viz. the regular conservative one since the counterpart of quasi-normality is forbidden by the flux conservation constraint. The experimental consequence is that leaky modes have been exploited for the investigation of dielectric structures over a long period of time (starting with the investigations by \Ron{Mie1908}), whereas irrefutable observational evidence of leaky MHD waves in the corona is still to come, see e.g.~\Ron{Wang2023}.

Actually, not only the normalisation of leaky MHD modes is a problem, but also the very concept of an infinite homogeneous plasma is a {\em contradictio in terminis.} Such plasmas do not exist in nature, not in the ordinary sense of finiteness of everything, but in the sense of not being a possible approximation of anything. All plasmas in space need currents or gravitational fields to confine them. The most obvious example is the tokamak experiment which could not contain hot plasmas without the enormous pressures exerted by the external current system. Astrophysical plasmas may not have visible coils but external currents are necessary anyway for equilibrium. On finite distances, the observational evidence of the corona is that it consists of myriads of loops continually interacting with each other. Consequently, multiple scattering of waves on ensembles of magnetic loops~\R{Keppens1994,KBG1994} is {\em the} tool for coronal seismology. It implies that an MHD wave emitted by a certain magnetic loop inevitably induces currents in an encountering loop, which reacts by reflecting part of the `outgoing' wave, which then is no longer exclusively outgoing. In other words, inhomogeneity is crucial to construct a reliable theory of coronal seismology to properly interpret the wide variety of observed MHD waves. One should also recall that for a proper theory of MHD spectroscopy, the assumption of a stationary {\em time-independent} equilibrium is necessary. This puts a limitation on the time scale of the waves, which should be  shorter than the time scale on which the surrounding loops react. If that condition cannot be met, full-blown nonlinear numerical  MHD is in order. Staying within the realm of linear MHD, keeping the one-dimensionality of an infinite cylinder, a first step in MHD spectroscopy would be to permit reasonable profiles for the different equilibrium quantities, both inside the primary loop and in the outer plasma, surrounded by a kind of `wall' at some average effective distance $R_{\rm eff}$ representing the scattering by all other loops. The pertinent boundary condition would be $\hs\xi_r(r = R_{\rm eff}) = 0\hs$. Next, if detailed data of the distribution of loops are available, azimuthal asymmetry in the angle $\vartheta$ would imply permitting a two-dimensional configuration with $R_{\rm eff} = f(\vartheta)$. The two-dimensional tools with finite element discretisation are presently readily available, the laborious task is to exploit them in a procedure that iteratively changes the equilibrium distributions such that the computed MHD spectra approach the observed ones. 

An other reason to consider wider structures of  coronal loops than the one-dimensional ones, is the effect of the photosphere on the MHD waves which is generally called {\em line-tying}. It is caused by the much higher density of the photosphere than that of the corona. For loops anchored in the photosphere on both ends $z = \pm \half L\hs$, it is frequently represented by the boundary conditions $\xi_r(r, z = \pm \half L) = 0$. However, this is not representative for line-tying, but for the {\em periodicity} of a finite cylinder approximation of toroidal plasmas like tokamaks. Line-tying demands that both components of the vector $\bfxi$ parallel to the photospheric interface vanish, i.e. $\hs\xi_r(r, z = \pm \half L) = \xi_\vartheta(r, z = \pm \half L)  = 0\hs$. Since those two components are out of phase, this implies that the line-tying boundary conditions cannot be satisfied by single harmonics $\exp ({\rm i}k \varphi)$. In effect, the problem has become intrinsically two-dimensional in $r$ and $\varphi$. It has been shown by \Ron{GoedHalber1994} that this implies a major change of the MHD spectrum drastically influencing even the Alfv\'en continuum frequencies. Subsequently, those were shown by~\Ron{Belien1996} to transform into a band structure with discrete Alfv\'en eigenmodes in the gaps. For completeness: once two-dimensionality is incorporated for the coordinates $r$ and $\vartheta$, it is obvious that the curvature of the loops should also be taken into account so that coronal seismology becomes a truly three-dimensional problem in $r$, $\vartheta$, $\varphi\hs$.

Finally, we need to discuss a counter argument to the conclusions of our previous paper \R{GKP2023}, and hence also applicable to the present one, brought forward by \Ron{Gao2024}. It is based on, e.g., investigations by~\Ron{Terradas2005} who treat the time-dependent initial value problem numerically for the standard plane slab model depicted in our Fig.~\F{1}. They find that the response to an initial impulse of a certain restricted spatial scale is, in fact, dominated by a short transient of damped leaky fast modes followed by a stationary state of bound fast-Alfv\'enic modes. The frequencies and damping rates are in perfect agreement with the theoretical predictions, hence also with the results shown in our Figs.~\F{7}--\F{10} of Sec.~\S{6}: the algebra is correct. An even nicer result is presented in a later paper by the same authors~\R{Terradas2006} for the cylindrical counterpart, in particular their Fig. 4 which again shows a response consisting of a short leaky mode transient but now followed by a damped quasi-mode due to the Alfv\'en continuum. It would be too cheap to discard these examples because they are theoretical calculations for a homogeneous outer medium. The point is that the initial response exhibits oscillations with a frequency and damping rate that exactly correspond to the computed leaky mode spectrum for this equilibrium. It is a reasonable assumption that such behaviour  would also be exhibited in reality by an observable loop during a limited period of time, provided (1)~that it would have {\em the same inhomogeneity} of the equilibrium fields, and (2)~that it would be excited with {\em the same impulse} with prescribed spatial and temporal cutoffs. 

The first pro viso concerns the methodology of coronal seismology. As mentioned above, its task is to solve the inverse spectral problem, i.e. the iterative determination of the possible distributions of the equilibrium fields of a certain loop that has produced an observed spectrum. This is a gigantic task, not even approximately started, witness the many sharp-boundary model calculations still in use in this field. Nevertheless, in principle, it can be carried out. The second proviso is a more serious one because it concerns the fundamentals of spectral theory. The initial perturbations that should be permitted in the solution of the inverse spectral problem are discussed in Sec.~\S{4.1}. This analysis gives the transparent solution~\E{4.15}: the response to an arbitrary initial perturbation $X(x;\omega_n)$ is given by the sum over all discrete exponentials multiplied by an integral over all space involving the initial perturbation. Here, the central problem of leaky modes is manifest. One should admit arbitrary initial data for an honest comparison of the responses in different configurations, but the integrals clearly diverge for leaky modes. This brings us back to the problem of normalisation discussed in Sec.~\S{4.2}. As discussed there, the oscillatory solutions for the leaky EM waves shown in Fig.~\F{6}, corresponding to the spectrum shown in Fig.~\F{4}, precisely cancel out the contributions of the two fields in the external region to produce satisfaction of the normalisation~\E{4.28} of the quasi-normal modes: systematic cutoff guaranteed! On the other hand, the oscillatory solutions for the leaky MHD waves shown in Fig.~\F{10}, corresponding to the spectrum shown in Fig.~\F{8}, do not cancel anything: there is no alternative for the regular normalisation. In that case, the initial impulse should be cutoff beyond a certain radius to prevent the exponentially increasing part of the solution to completely dominate the solution. The problem is: what cutoff radius to take? Its value will determine which part of the leaky mode spectrum will participate in the truncated problem, and it will be different for the various equilibria and associated MHD spectra investigated. Consequently, cutoff criteria are completely arbitrary so that there is no guiding principle to interpret initial transients, and a systematic use of leaky modes for coronal seismology is illusory.

\ver{2}

In summary, for all the reasons given here, we maintain the final conclusion of our previous paper \R{GKP2023}: {\em Exit leaky modes in coronal magnetic flux tubes!}
To which we now add: {\em In particular, in solar seismology.}

\ver{10}


\noindent
{\bf Declaration of interest} The authors report no conflict of interest.

\acknowledgements
\noindent
{\bf Acknowledgements} DIFFER is part of the institutes organisation of NWO. 
This research was supported by the KU Leuven (GOA/2015-014).

\bibliographystyle{jpp}
\bibliography{OM2026}

\vfill\eject 

\begin{figure}
 \centerline{\includegraphics[width=0.5\textwidth]{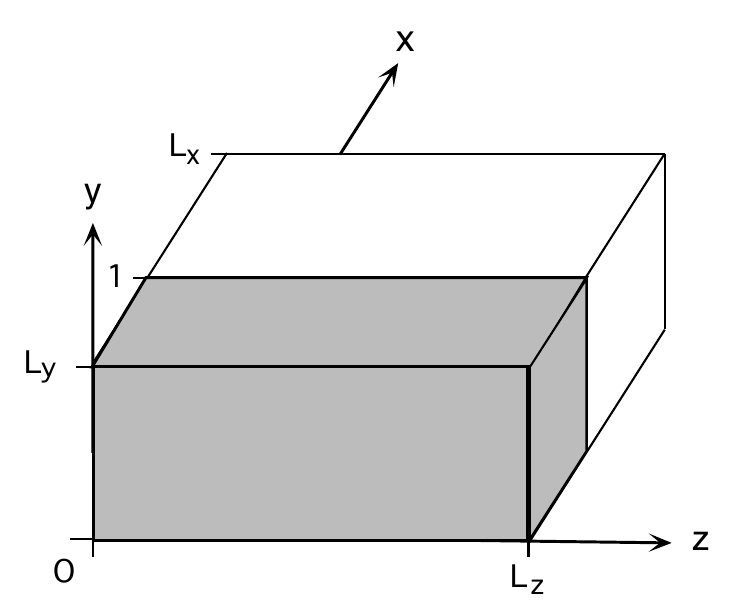}}
  \caption{Schematic geometry for a dielectric slab, or a coronal flux tube, infinitely extended 
in all directions ($L_x , L_y, L_z \rightarrow \infty$). The phase velocity $v(x)$ varies inside the 
slab ($|x| \le 1$), possibly with only a jump at $|x| = 1$. Elsewhere, the phase velocity is 
constant (equal to the velocity of light in vacuum). In the $x$-direction, the waves are imposed 
to be symmetric (even) or anti-symmetric (odd) with respect to the plane $x = 0$. Extensions 
along the negative directions are suppressed. All lengths are made dimensionless in units of 
the semi-thickness $a$ of the slab.}{\Fn{1}}
\end{figure}

\begin{figure}
 \centerline{\includegraphics[width=0.6\textwidth]{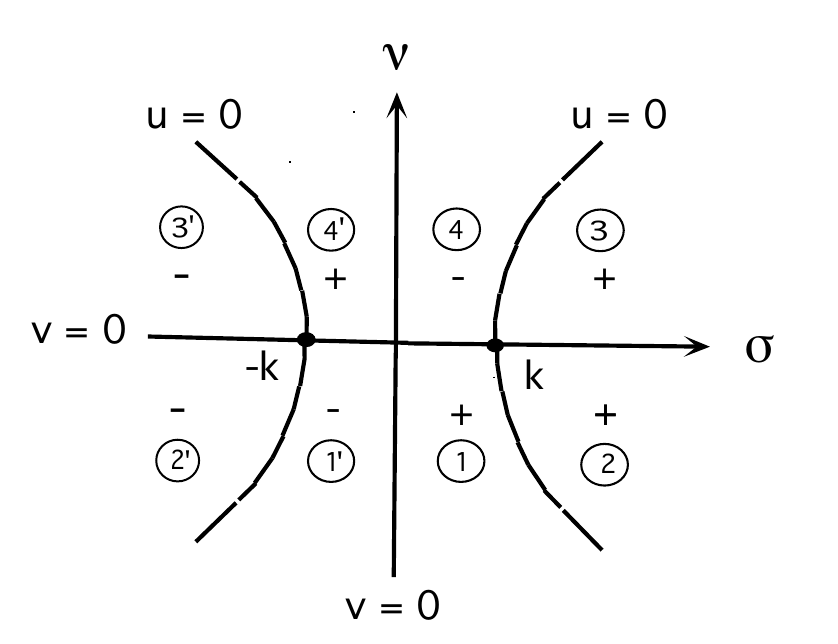}}
  \caption{The complex $\omega$-plane is divided by the hyperbola $u = 0$, i.e\ $\sigma^2 - 
\nu^2 = k^2$, and the lines $v = 0$, i.e\ $\sigma = 0$ and $\nu = 0$, in eight regions with the 
indicated choice of the $\pm$ sign in the expression~\E{2.14} for the external wavenumber 
$\hat{q}$. This implies
\newline 
$\hor{5}$ -- for region $\bigcirc{\hor{-2.3}{\scriptstyle 1}}\;\,,\;  u \le 0 \,,\; \nu \le 0 \;$:  
\quad $- \half \pi \le \varphi <  - \onefourth \pi \quad\Rightarrow\quad \hat{q}_1 \ge 0 \,, \, 
\hat{q}_2 < 0\hs$;
\newline
$\hor{5}$ -- for region $\bigcirc{\hor{-2.3}{\scriptstyle 2}}\;\,,\; u > 0 \,,\; \nu \le 0 \;$: \quad 
$- \onefourth \pi < \varphi \le  0 \hor{5}\quad\Rightarrow\quad \hat{q}_1 > 0 \,, \, \hat{q}_2 
\le 0\hs$;
\newline 
$\hor{5}$ -- for region $\bigcirc{\hor{-2.3}{\scriptstyle 3}}\;\,,\; u > 0 \,,\; \nu > 0 \;$: \quad 
$\hor{5}0 < \varphi < \onefourth \pi \hor{3}\quad\Rightarrow\quad \hat{q}_1 > 0 \,, \, 
\hat{q}_2 > 0 \hs$; 
\newline 
$\hor{5}$ -- for region $\bigcirc{\hor{-2.3}{\scriptstyle 4}}\;\,,\; u \le 0 \,,\; \nu > 0 \;$:  \quad 
$\hor{3}\onefourth \pi \le \varphi <  \half \pi \hor{3}\quad\Rightarrow\quad \hat{q}_1 < 0 \,, 
\, \hat{q}_2 < 0\hs$.
\newline
The sign changes across the real axis are crucial to compute the Spectral Webs of
Fig.~\F{4}.}{\Fn{2}}
\end{figure}

\vfill\eject 

\begin{figure}
 \centerline{\includegraphics[width=0.9\textwidth]{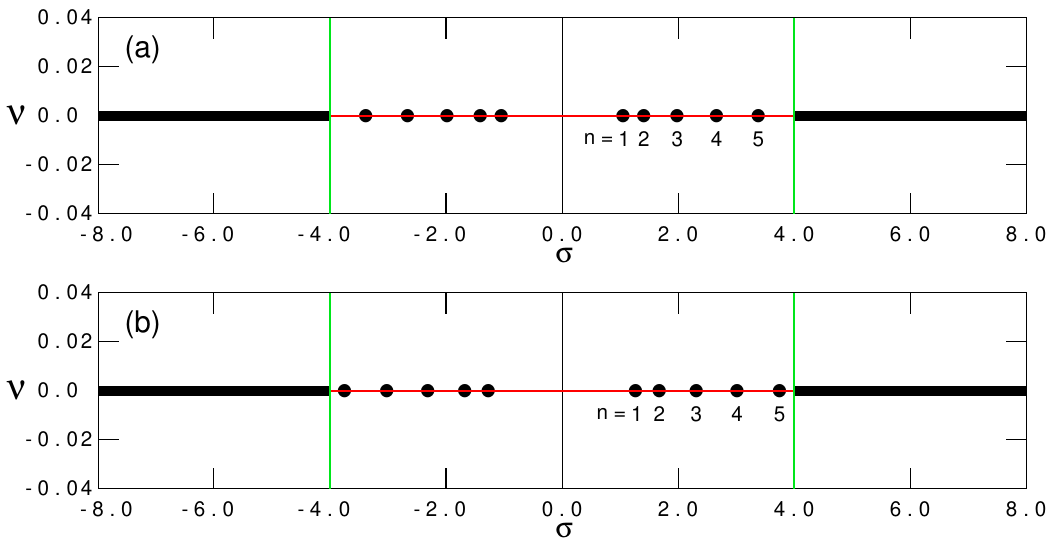}}
  \caption{Spectra of Conservative Normal Modes, i.e.\ regular TE waves in the 
dielectric medium depicted in Fig.~\F{1}: (a)~even modes, (b)~odd modes. The 
phase velocity $v(x)$ changes rapidly from $\delta = 0.25$ to unity in a narrow 
region of size $\epsilon = 0.02$ just below $|x| = 1$. The wave number in the 
longitudinal ($z$)-direction is $k = 4.0$. The mode numbering $n = 1 \dots 5$ 
refers to the bound modes, the `propagating' (standing) waves for $|\sigma| \ge k$ constitute continuous 
spectra.}{\Fn{3}}
\end{figure}

\begin{figure}
 \centerline{\includegraphics[width=0.9\textwidth]{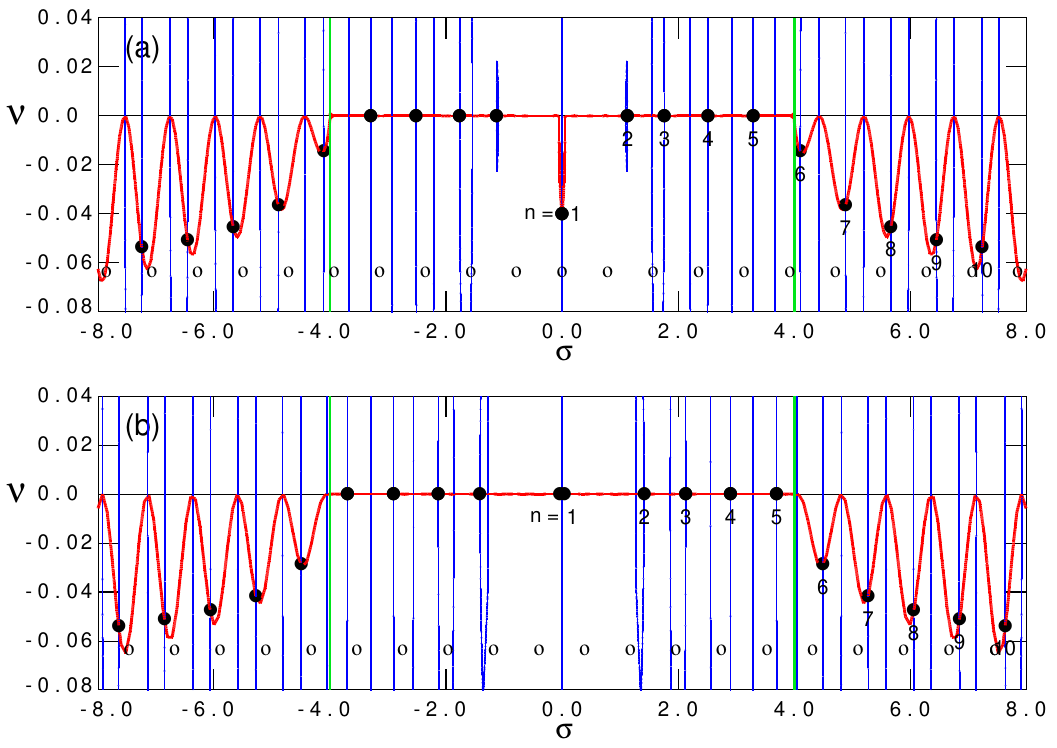}}
  \caption{Spectra of Quasi-Normal Modes, i.e.\ leaky TE modes in the dielectric 
medium depicted in Fig.~\F{1}: (a)~even modes, (b)~odd modes. The parameters 
$\delta$, $\epsilon$ and $k$ are the same as in Fig.~\F{3}. The mode numbering 
$n = 1 \dots 5$ refers to the unbound (cutoff) discrete modes. Mode numbering of the discrete 
complex (propagating) modes for $|\sigma| \ge k$ may be continued with $n = 6 \dots \infty$ by exploiting monotonicity along the solution
path (red) of the Spectral Web separated by the intersections with the conjugate path (blue). Intersections also appear on the real axis, for $S_{\rm com} = 0$, 
but those `solutions' are spurious since $E(1) = 0$ there. The open circles correspond to the standard 
step function model with $k = 0$, where $\nu = \half \delta \ln[(1 - \delta)/(1 + \delta)]$.}{\Fn{4}}
\end{figure}

\vfill\eject 

\begin{figure}
 \centerline{\includegraphics[width=0.65\textwidth]{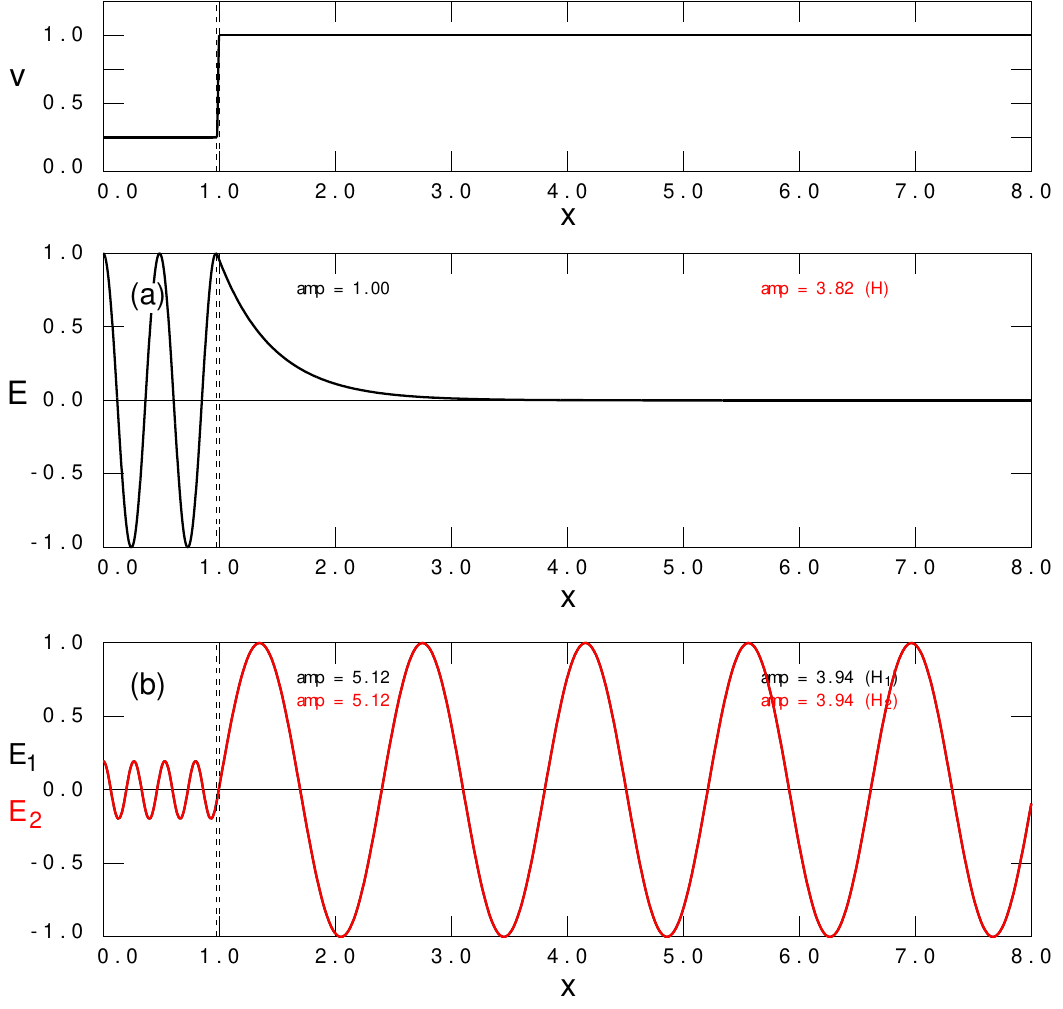}}
  \caption{CNM eigenfunctions corresponding to the spectrum depicted in 
Fig.~\F{3}(a) for the even modes: (a)~a discrete bound mode with $n = 5$ and 
$\sigma = 3.3813$ on the interval $(0, 8.0)$; (b)~a `propagating' continuum 
mode with the arbitrary choice $\sigma = 6.0$ of the improper eigenvalue on the 
same interval. The phase velocity profile $v(x)$  for this 
interval is depicted above the eigenfunctions. The parameters $\delta$, 
$\epsilon$ and $k$ are the same as in Fig.~\F{3}. The amplitudes of the associated functions $H_{1,2}(x)$, which are not displayed, {\em decrease} across the inhomogeneous boundary layer (in contrast to the functions~$E_{1,2}(x)$ which increase). Their amplitudes at $|x| = 1$ are given in the top right of the figures.}{\Fn{5}}
\end{figure}

\begin{figure}
 \centerline{\includegraphics[width=0.65\textwidth]{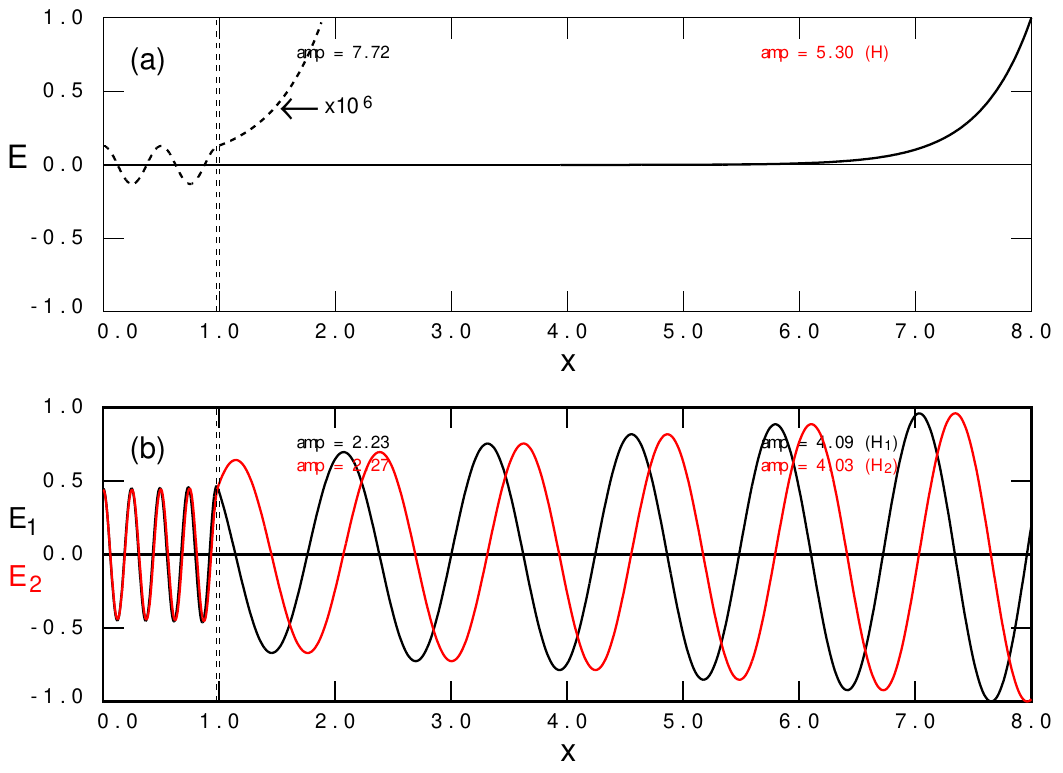}}
  \caption{QNM eigenfunctions corresponding to the spectrum depicted in 
Fig.~\F{4}(a) for the even modes: (a)~A real, unbound, discrete mode with $n = 
4$ and $\sigma = 3.2983$ on the interval $(0, 8.0)$. Since the exponential 
growth is huge, the eigenfunction has been magnified by the factor $10^6$ 
(dashed curves) to show the oscillations on the internal part of the interval. 
(b)~A complex, leaky, discrete mode with $n = 9$ and $\sigma = 6.4545$, $\nu = -
0.050836$ on the same interval. The parameters $\delta$, 
$\epsilon$ and $k$ are the same as in Fig.~\F{3}.}{\Fn{6}}
\end{figure}

\vfill\eject 

\begin{table}
\centerline{\includegraphics[width=0.75\textwidth]{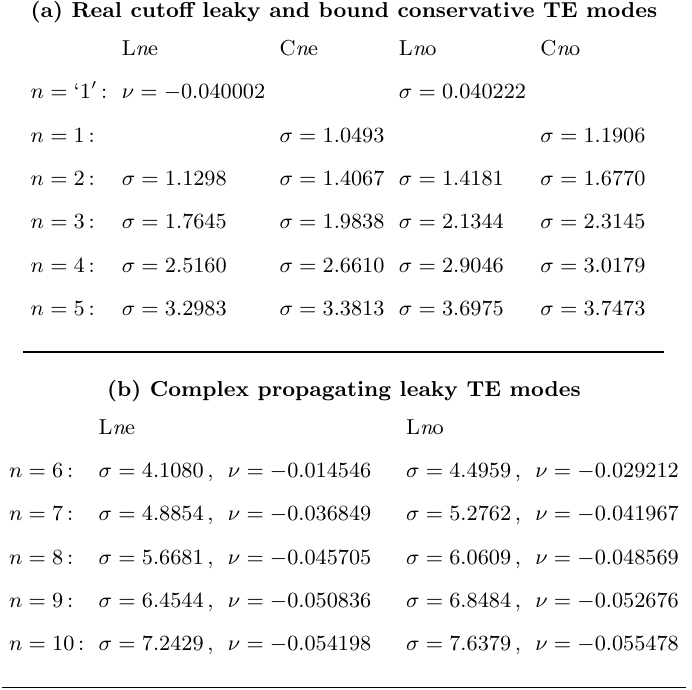}}
\caption{Connection between the different modes of the spectra depicted in Figs.~\F{3} and \F{4}:\newline
(a)~Alternating real frequencies of the even leaky (${\rm L}{n{\rm e}}$),
the even conservative (${\rm C}{n{\rm e}}$),
the odd leaky (${\rm L}{n{\rm o}}$),
and the odd conservative (${\rm C}{n{\rm o}}$) modes in the cutoff region. The first two leaky modes ($n = `1'$) immediately start to diverge away from $|x| = 0$ so that they strictly fall outside the alternating $n > 1$ sequences, which are based on monotonicity along the solution path.\newline
(b)~Complex frequencies of the even (${\rm L}{n{\rm e}}$) and odd(${\rm L}{n{\rm o}}$) propagating leaky modes. Beyond cutoff, the real frequencies of the even/odd conservative modes form continua of standing waves.}{\Tn{1}}
\end{table}

\vfill\eject 

\begin{figure}
 \centerline{\includegraphics[width=0.78\textwidth]{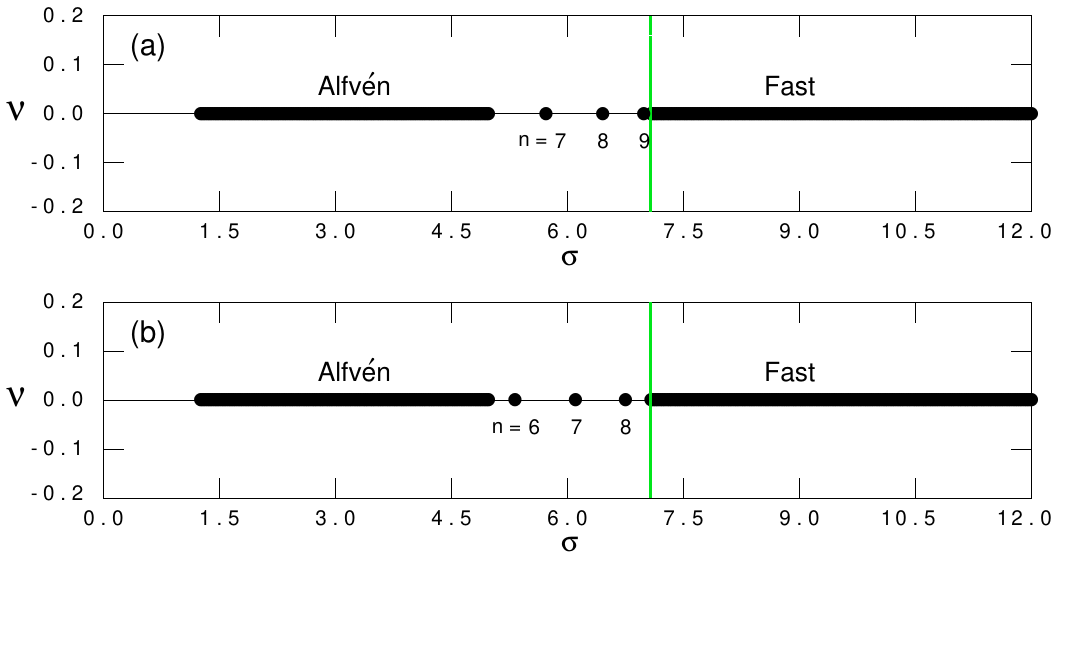}}
  \caption{Spectra of Conservative Normal Modes, i.e. regular MHD waves in a 
coronal magnetic flux tube as depicted in Fig.~\F{1}: (a)~even modes, (b)~odd modes. 
The parameters are: $\epsilon = 0.2$, $\delta = 0.25$, $k = \ell = 5.0$. 
The modes $n = 1 \cdots 6$, or $n = 1 \ldots 5$, have been `swallowed' by the Alfv\'en continua.
The remaining modes $n = 7, 8, 9$, or $n = 6, 7, 8$, 
refer to discrete bound modes which for $|\sigma| \ge k_{\rm tot}$ continuously connect to the continua 
 of finite amplitude standing fast waves.}{\Fn{7}}
\end{figure}

\begin{figure}
 \centerline{\includegraphics[width=0.78\textwidth]{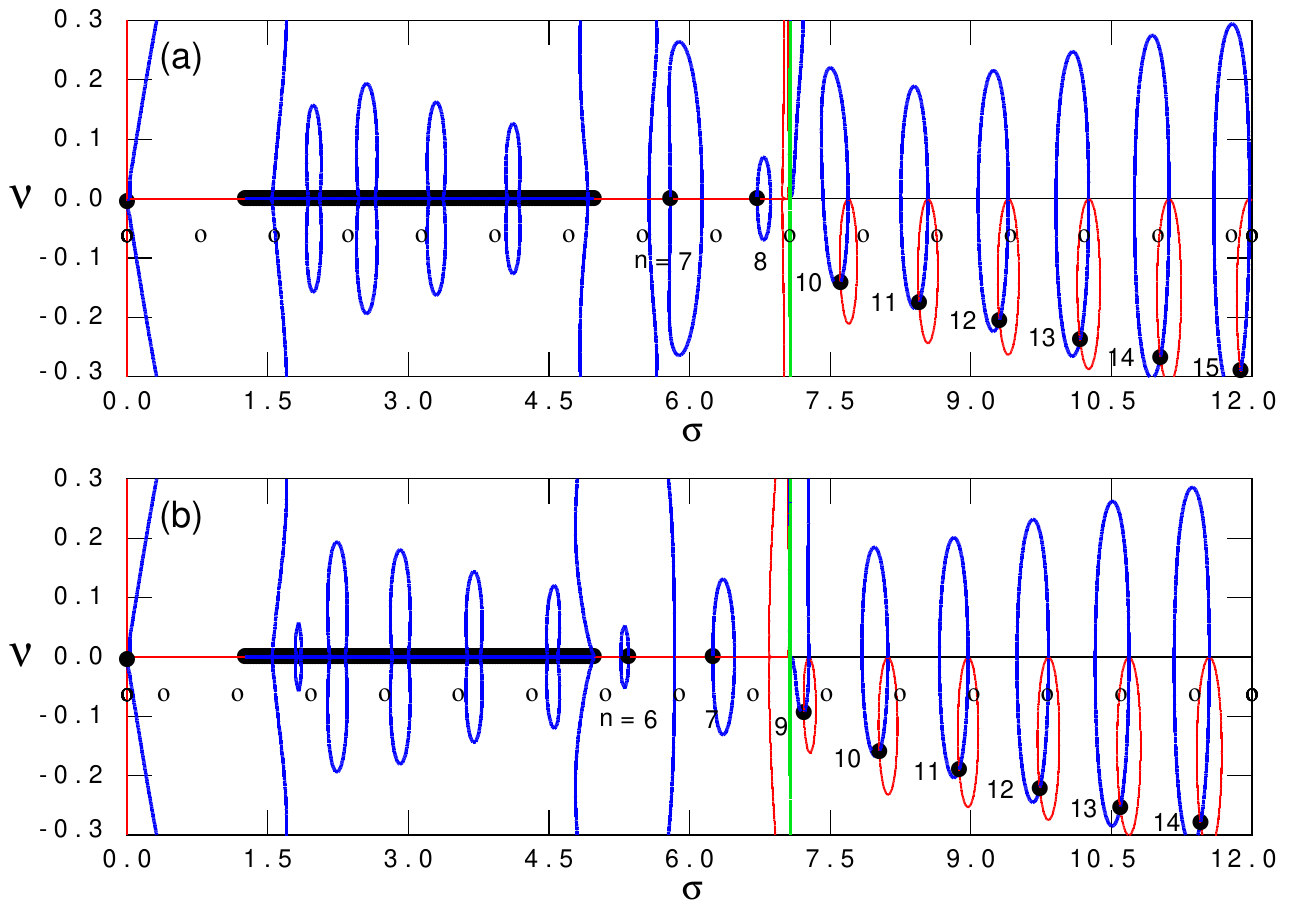}}
  \caption{Spectra of leaky MHD waves in a 
coronal magnetic flux tube as depicted in Fig.~\F{1}: (a)~even modes, (b)~odd modes. The parameters 
 are the same as in Fig.~\F{7}. Again, the lower modes have been `swallowed' by the Alfv\'en continua.
 The remaining modes $n = 7, 8$, or $n = 6, 7$ refer to discrete unbound modes which for
  $|\sigma| \ge k_{\rm tot}$ discontinuously `connect' onto the discrete spectra of propagating fast waves with 
  exponentially growing amplitudes for $|x| \rightarrow \infty$.}{\Fn{8}}
\end{figure}

\vfill\eject 

\begin{figure}
 \centerline{\includegraphics[width=0.7\textwidth]{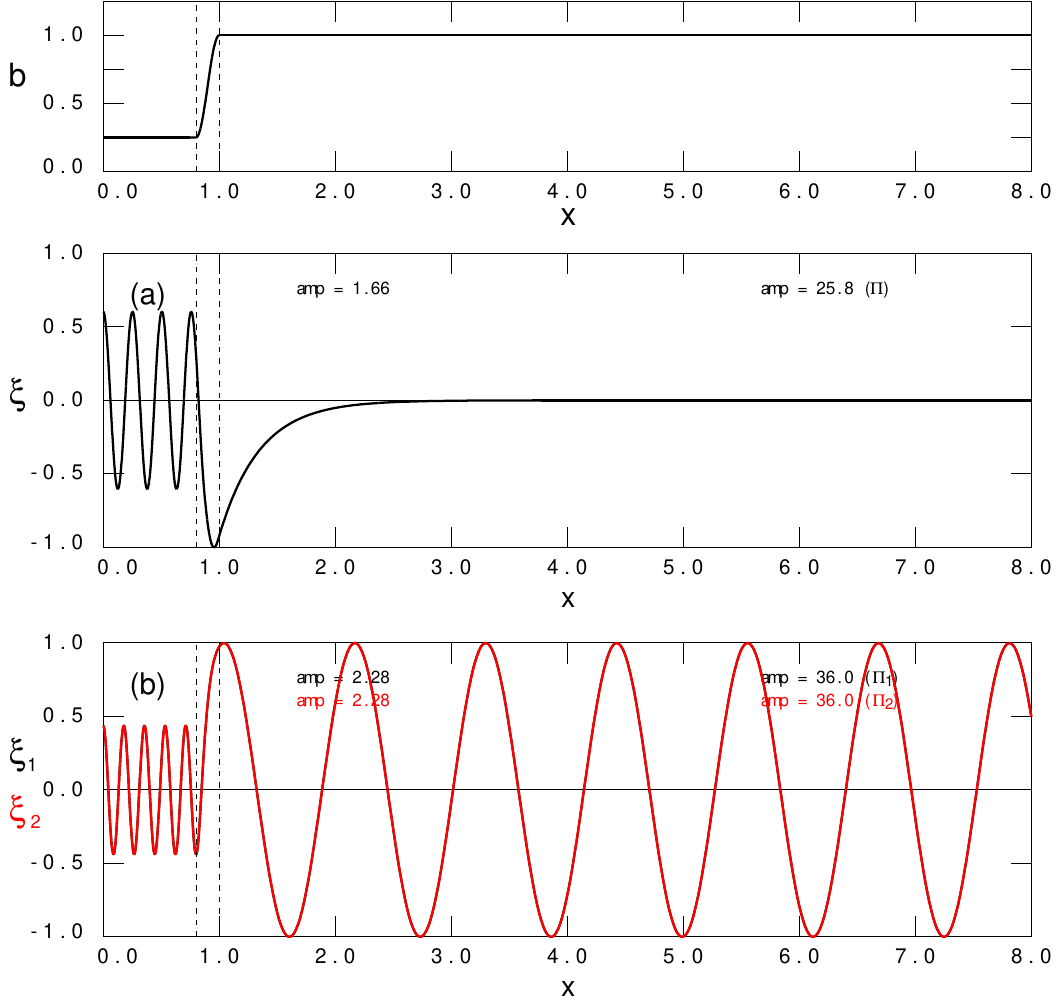}}
  \caption{CNM eigenfunctions corresponding to the spectrum depicted in 
Fig.~\F{7}(a) for the even modes: (a)~a discrete bound mode with $n = 8$ and 
$\sigma = 6.4563$ on the  interval $(0, 8.0)$; (b)~a `propagating' continuum 
mode with the arbitrary choice $\sigma = 9.0$ of the improper eigenvalue on the 
same interval. The phase velocity profile $b(x)$  for this 
interval is depicted above the eigenfunctions. The parameters $\delta$, 
$\epsilon$ and $k$ are the same as in Fig.~\F{7}.}{\Fn{9}}
\end{figure}

\begin{figure}
 \centerline{\includegraphics[width=0.7\textwidth]{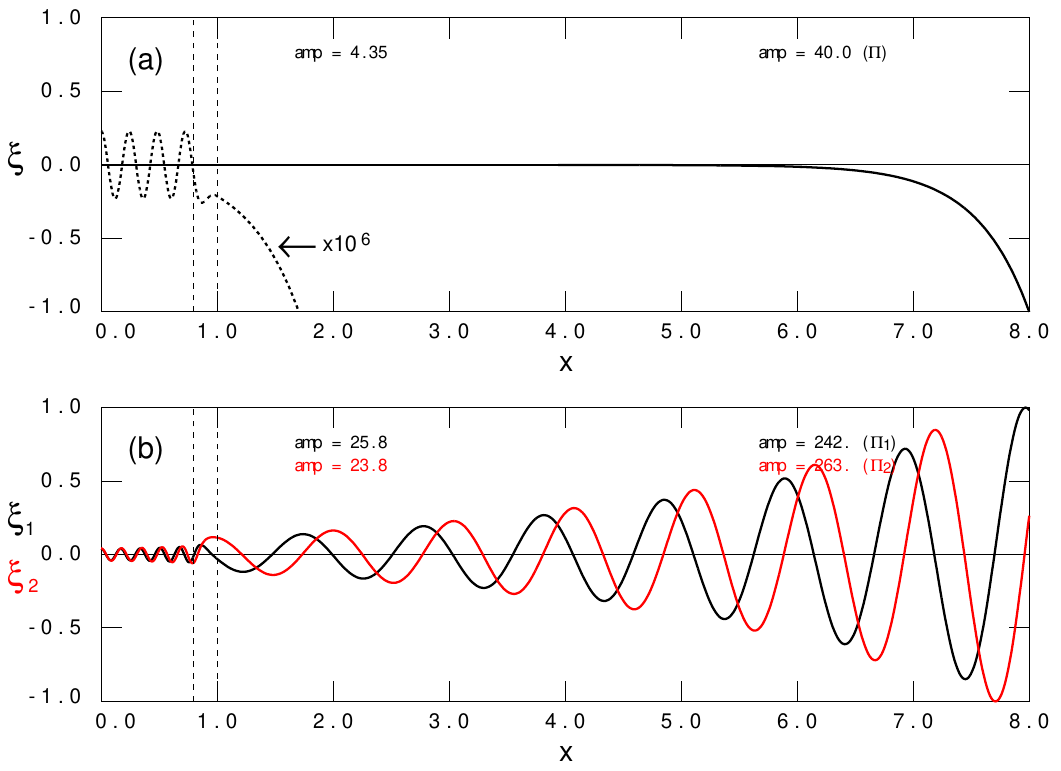}}
  \caption{Leaky `eigenfunctions' corresponding to the spectrum depicted in 
Fig.~\F{4}(a) for the even modes: (a)~A real, unbound, discrete mode with $n = 
8$ and $\sigma = 6.7219$ on the interval $(0, 8.0)$. Since the exponential 
growth is huge, the eigenfunction has been magnified by the factor $10^6$ 
(dashed curves) to show the oscillations on the internal part of the interval. 
(b)~A complex, leaky, discrete mode with $n = 12$ and $\sigma = 9.3059$, $\nu = -
0.20632$ on the same interval. The parameters $\delta$, 
$\epsilon$ and $k$ are the same as in Fig.~\F{7}.}{\Fn{10}}
\end{figure}

\vfill\eject 

\end{document}